\documentclass[oldversion]{aa}  
\usepackage{graphicx}
\usepackage{txfonts}

\def\0{\phantom0}

\def\kms{km s$^{-1}$}
\def\obj{QSO~2237$+$0305}

\begin{document}

\title{Microlensing variability in the gravitationally lensed quasar \\    
\vspace*{1mm}  \obj\ $\equiv$ the Einstein Cross
 \thanks{Based on observations made 
 with the ESO-VLT Unit Telescope \#~2 Kueyen 
 (Cerro Paranal, Chile; Proposals 
 073.B-0243(A\&B),
 074.B-0270(A), 
 075.B-0350(A), 
 076.B-0197(A),
 177.B-0615(A\&B), PI: F. Courbin).}
 }
\subtitle{I. Spectrophotometric monitoring with the VLT}

\author{
A. Eigenbrod\inst{1} \and F. Courbin\inst{1} \and  D. Sluse\inst{1}
\and G. Meylan\inst{1}
\and E. Agol\inst{2} }


\institute{
Laboratoire d'Astrophysique, Ecole Polytechnique F\'ed\'erale
de Lausanne (EPFL), Observatoire de Sauverny, 1290 Versoix, Switzerland
\and
Astronomy Department, University of Washington, 
Box 351580, Seattle, WA 98195, USA
}

\date{Received ... ; accepted ...}
\authorrunning{A. Eigenbrod et al.}
\titlerunning{Microlensing variability in the Einstein Cross}

\abstract{We present the results of the first long-term (2.2 years) spectroscopic monitoring
of a gravitationally lensed quasar, namely the Einstein Cross \obj. 
The goal of this paper is to present the
observational facts to be compared in follow-up papers with theoretical models 
to constrain the inner structure of the source quasar.

We spatially deconvolve deep VLT/FORS1 spectra 
to accurately separate the spectrum of the  lensing galaxy from the
spectra of the quasar images. Accurate cross-calibration of the 
58 observations at 31-epoch
from October 2004 to December 2006
is carried out with non-variable foreground stars observed simultaneously with the 
quasar.  The quasar spectra are further 
decomposed into a continuum component and several broad emission lines
to infer the variations of these spectral components.

We find prominent microlensing events in the quasar images A and B,
while images C and D are almost quiescent on a timescale of a few months. 
The strongest variations are observed in the continuum of image A. Their amplitude is larger in the
blue (0.7~mag) than in the red (0.5~mag), consistent with microlensing of an accretion disk. Variations 
in the intensity and profile of the broad emission lines are also reported, 
most prominently
in the wings of the \ion{C}{III]} and center of the \ion{C}{IV} emission lines. 
During a strong microlensing episode observed in June 2006  in quasar image A, the broad 
component of the \ion{C}{III]} is more highly magnified than the narrow component. 
In addition, the emission lines with higher ionization potentials are 
more magnified than the lines with lower ionization potentials, consistent with 
the results obtained with reverberation mapping. Finally, we find that the V-band differential
extinction by the lens, between the quasar images, is in the range $0.1-0.3$~mag.
}

\keywords{Gravitational lensing: quasar, microlensing --- 
    Quasars: general.
    Quasars: individual QSO~2237$+$0305, Einstein Cross}

\maketitle

\section{Introduction}

The gravitational lens  \obj, also known as  ``Huchra's lens''  or the
``Einstein  Cross'', was discovered  by Huchra et al.  (\cite{huchra})
during the Center for Astrophysics  Redshift Survey.  It consists of a
$z_s=1.695$ quasar gravitationally lensed into four images arranged in
a crosslike  pattern around the  nucleus of a $z_l=0.0394$  barred Sab
galaxy. The average projected distance of the images from the lens center is 
$0.9$\arcsec. 

A few years after  this discovery, Schneider et al. (\cite{schneider})
and  Kent \& Falco (\cite{kent})  computed  the first simple models of
the  system,  leading to the  conclusion   that this  system was  very
promising to study microlensing.    Indeed, the predicted  time delays
between the four quasar images are of the order of a day 
(Rix et al. \cite{rix92}, Wambsganss \& Paczy\'nski \cite{wambsganss94}), 
meaning that
intrinsic  variability of the  quasar can easily be distinguished from
microlensing events. In  addition, the particularly small  redshift of
the lensing galaxy    implies  large tangential velocities  for    the
microlenses. Furthermore the quasar images form right  
in  the bulge  of  the lens where the stellar  density is the highest.  
The   combination of these
properties makes microlensing events very likely  in the Einstein Cross
and   very rapid,  with   timescales   of a   few   weeks to  a   few
months. Indeed,  Irwin    et al.   (\cite{irwin})  reported  significant
brightness variations of the  brightest   quasar image A,  which  they
interpreted as the first detection  ever of microlensing in the images
of a multiply-imaged quasar.

Since then, microlensing events have been observed in several other gravitationally lensed
quasars, and are  expected to occur  in virtually  any quadruply lensed
quasar (Witt et   al.   \cite{witt}).  Probably the   most  compelling
examples of   microlensing  light curves   are given   by  the Optical
Gravitational       Lensing   Experiment     (OGLE)   (Wo\'zniak    et
al.  \cite{wozniak00a}, Udalski et al. \cite{udalski}).  
Started in 1997,  this project monitors regularly
the four  quasar images of \obj,   showing continuous
microlensing-induced variations in the light curves.

Most of  the quasar  microlensing studies so  far  are  based  exclusively on
broad-band  photometric monitoring    (e.g.  Wo\'zniak  et   al.
\cite{wozniak00b};   Schechter et   al.   \cite{schechter}; Colley  \&
Schild \cite{colley}). These  observations,  even though  dominated by
variations   of the continuum, make   it very difficult to disentangle
variations  in the  continuum  from variations in the  broad emission lines
(BELs). Both types of regions are affected by microlensing, but in 
different ways depending on their size.

Microlensing of an extended source can  occur when its size is smaller
than or comparable to  the  Einstein radius of  a  star, i.e. of  the
order of $10^{17}$~cm or $10^{-1}$~pc in the case of the Einstein
Cross (Nemiroff \cite{nemiroff}; Schneider \& Wambsganss
\cite{sw90}).  From reverberation mapping, the broad line region (BLR) 
was long estimated to be larger  than this, of the order
of   $10^{18}$~cm or  $1$~pc, hence     leaving  little room  for  BEL
microlensing.  However,  more  recent  reverberation  mapping  studies
revise this downwards, to $10^{16}$~cm (Wandel et al.
\cite{wandel}, Kaspi et al.  \cite{kaspi}), which is  also
consistent with the disk-wind model of Murray et al.  (\cite{murray}).
Inspired by these numbers, Abajas et~al.  (\cite{abajas}) and Lewis \&
Ibata (\cite{lewis})  investigated BEL microlensing  in further  detail
and computed possible line profile variations for various BLR models.

Observations  of significant continuum and  BEL microlensing have been
reported in a number of systems 
(QSO~2237$+$0305, Filippenko \cite{filippenko}, Lewis et al. \cite{lewis98}, 
                 Wayth et al. \cite{wayth};
HE~2149$-$2745, Burud et al. \cite{burud}; 
HE~0435$-$1223, Wisotzki et al. \cite{wisotzki03};
H~1413$+$117, Chartas et al. \cite{chartas};
SDSS~J1004$+$4112, Richards et al. \cite{richards};   
HE~1104$-$1805, G\'omez-\'Alvarez \cite{gomez04};
HE~0047$-$1756, Wisotzki et al. \cite{wisotzki04};
SDSS~J0924$+$0219, Eigenbrod et al. \cite{cosmograil2}, Keeton et al. \cite{keeton05} ;
and RXJ~1131$-$1231, Sluse et al. \cite{sluse}).
These first  observational indications of  microlensing  can be turned
into a powerful  tool to  probe the  inner parts of  quasars, provided
regular spectroscopic data can be obtained.  Several theoretical studies
show how multiwavelength light curves can constrain the energy profile of
the quasar   accretion disk and also  the  absolute sizes  of the 
line-emitting regions (e.g., Agol \& Krolik~\cite{agol}, Mineshige \& Yonehara
\cite{mineshige}, Abajas et al.~\cite{abajas},  Kochanek \cite{kochanek}). 

In  this   paper we   present the   results  of   the  first long-term
spectrophotometric monitoring  of  the Einstein  Cross.   The
spectral variations  of the  four   quasar images  are followed  under
sub-arcsecond seeing conditions with  the Very
Large   Telescope (VLT) of   the  European Southern  Observatory
  for more than two  years,
from October 2004 to December 2006, with a mean
temporal sampling  of about one point every   second week.  This first
paper describes the observations, the method used to separate the
quasar spectra   from  that of   the lensing   galaxy,  and  the  main
observational results. Simple considerations of the properties of microlensing caustics 
and of the geometry of the central parts of the quasar already allow us to infer
interesting constraints on the quasar energy profile in the Einstein Cross.

The full analysis of our monitoring data, still  being acquired at the
VLT,  requires detailed  microlensing  simulations coupled with  quasar
models and will be the subject of future papers.

\begin{figure}[ht!]
\begin{center}
\includegraphics[width=8.5cm]{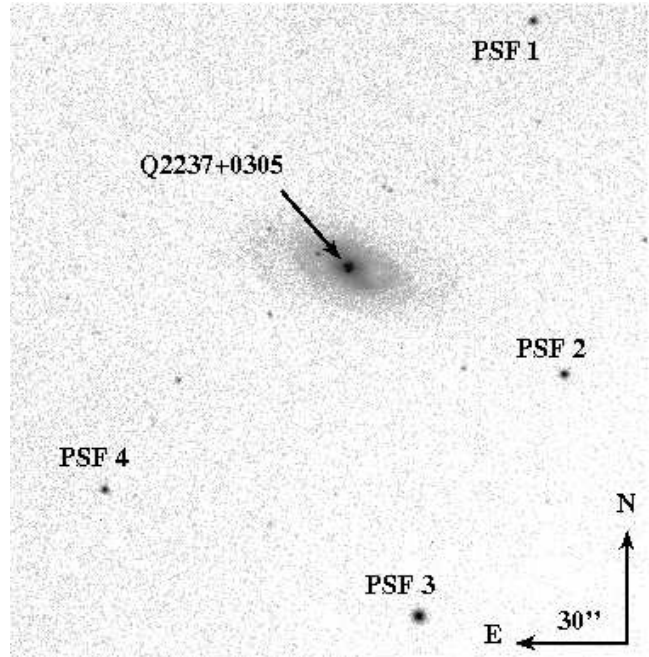}
\caption{VLT/FORS1 field of view showing the lensed quasar \obj, along with
the four PSF  stars used to spatially deconvolve  the spectra.   These
stars  are also used to  cross-calibrate the observations in flux from
one epoch to another and to minimize the effect of sky transparency.}
\label{field}
\end{center}
\end{figure}

\section{Observations}

We acquired our   observations   with   the FOcal Reducer   and low
dispersion     Spectrograph (FORS1), mounted on Kueyen, 
the Unit Telescope \#~2  of the ESO Very Large
Telescope (VLT) located at Cerro Paranal (Chile).  We performed our observations
in the  multi-object spectroscopy (MOS) mode.   This
strategy allowed us to get simultaneous observations of the main target
and   of  four  stars    used  as   reference point-spread   functions
(PSFs). These stars  were used to  spatially deconvolve the  spectra, as
well  as to perform accurate flux  calibration of the target spectra
from one epoch to another.   We  chose these stars   to be located  as
close  as  possible  to   \obj\ and    to   have  similar  apparent
magnitudes as the quasar images. 
The PSF stars \#~1, 2, 3, and 4 have R-band magnitudes of 17.5, 17.0, 
15.5, and 17.5~mag, respectively.
Fig.~\ref{field} shows the field of view of our observations.

We used the high-resolution collimator of FORS1 to achieve the
best possible spatial sampling   of the data, i.e.  $0.1\arcsec$  per
pixel.  With this  resolution, we observed a maximum  of 8 objects
simultaneously      over   a         field         of  view        of
$3.4\arcmin\times3.4\arcmin$. One slit was  aligned  along two  of  the
quasar images and four slits were centered on foreground PSF stars. We
placed the
remaining slits on empty sky regions  and used them to carry
out sky subtraction  of the  quasar  data.

\begin{figure}[t!]
\begin{center}
\includegraphics[width=8.5cm]{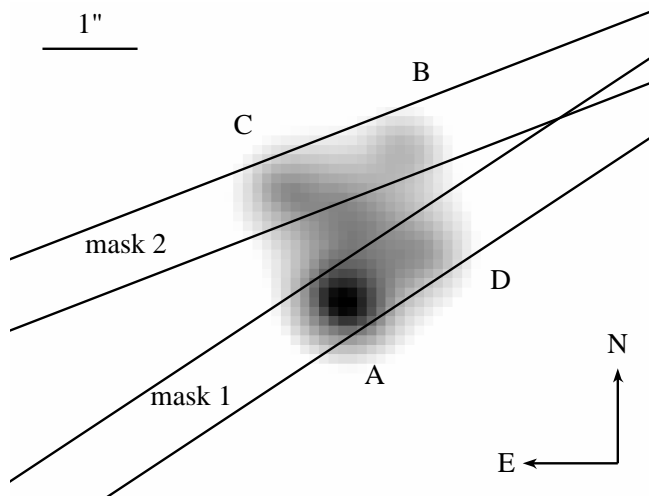}
\caption{FORS1 R-band acquisition image of QSO~2237$+$0305 taken on epoch \#~12
(12$-$09$-$2005). The slits used in the two masks are shown. Their width is
$0.7$\arcsec. The Position Angle (PA) of  mask~1 is $PA=+56.5^{\circ}$ and that 
of mask~2 is $PA=+68.5^{\circ}$.}
\label{Q2237_slits}
\end{center}
\end{figure}

Two masks were  designed to observe the two pairs of quasar images. The PSF stars in both
masks were the same. Fig.~\ref{Q2237_slits}  shows the slit positioning
with respect to  our target. 
The first mask was aligned on quasar images A and D, while the second
was aligned on images B and C. The masks  were
rotated to position angles that  avoid clipping of any quasar image.
This  is mandatory to carry out  spatial deconvolution of the spectra. 

Our  observing sequences consisted of    a short acquisition image,   an
``image-through-slit'' check,    followed   by  a   consecutive   deep
spectroscopic exposure.  All individual  exposures were $1620$~s  long.
We list the journal of our observations in Table~\ref{journal}. The mean
seeing during the three observing seasons was $0.8$\arcsec. We chose a slit 
width of $0.7$\arcsec, approximately matching the seeing and much smaller than the
mean separation  of $1.4$\arcsec between the  quasar images. 
This is mandatory to avoid  contamination of an image by the others.

\begin{table}[p]
\caption[]{Journal of the observations taken on 31 epochs.}
\label{journal}
\begin{flushleft}
\begin{tabular}{clcccc}
\hline 
\hline 
ID & Civil Date & HJD & Mask & Seeing $[\arcsec]$ & Airmass \\
\hline  					       
1  & 13$-$10$-$2004& 3292 & 1 & 0.86 & 1.204 \\    
1  & 14$-$10$-$2004& 3293 & 2 & 0.87 & 1.221 \\    
2  & 14$-$11$-$2004& 3324 & 1 & 0.75 & 1.184 \\    
2  & 14$-$11$-$2004& 3324 & 2 & 0.68 & 1.305 \\    
3  & 01$-$12$-$2004& 3341 & 1 & 0.88 & 1.355 \\    
3  & 01$-$12$-$2004& 3341 & 2 & 0.94 & 1.609 \\    
4  & 15$-$12$-$2004& 3355 & 1 & 0.99 & 1.712 \\    
4  & 16$-$12$-$2004& 3356 & 2 & 0.90 & 1.817 \\    
5  & 11$-$05$-$2005& 3502 & 1 & 0.87 & 1.568 \\    
5  & 12$-$05$-$2005& 3503 & 2 & 0.51 & 1.389 \\    
6  & 01$-$06$-$2005& 3523 & 1 & 0.63 & 1.342 \\    
6  & 01$-$06$-$2005& 3523 & 2 & 0.64 & 1.224 \\    
7  & 01$-$07$-$2005& 3553 & 1 & 0.57 & 1.153 \\    
8  & 14$-$07$-$2005& 3566 & 1 & 0.89 & 1.620 \\    
9  & 06$-$08$-$2005& 3589 & 1 & 0.51 & 1.135 \\    
9  & 06$-$08$-$2005& 3589 & 2 & 0.61 & 1.173 \\    
10 & 15$-$08$-$2005& 3598 & 1 & 0.86 & 1.140 \\    
11 & 25$-$08$-$2005& 3608 & 1 & 0.49 & 1.261 \\    
11 & 25$-$08$-$2005& 3608 & 2 & 0.54 & 1.461 \\    
12 & 12$-$09$-$2005& 3626 & 1 & 0.70 & 1.535 \\    
12 & 12$-$09$-$2005& 3626 & 2 & 0.69 & 1.341 \\    
13 & 27$-$09$-$2005& 3641 & 1 & 0.92 & 1.480 \\    
13 & 27$-$09$-$2005& 3641 & 2 & 0.73 & 1.281 \\    
14 & 01$-$10$-$2005& 3645 & 1 & 0.78 & 1.281 \\    
14 & 01$-$10$-$2005& 3645 & 2 & 0.87 & 1.156 \\    
15 & 11$-$10$-$2005& 3655 & 1 & 0.57 & 1.140 \\    
15 & 11$-$10$-$2005& 3655 & 2 & 0.66 & 1.134 \\    
16 & 21$-$10$-$2005& 3665 & 1 & 0.70 & 1.215 \\    
16 & 21$-$10$-$2005& 3665 & 2 & 0.74 & 1.156 \\    
17 & 11$-$11$-$2005& 3686 & 1 & 0.90 & 1.137 \\    
17 & 11$-$11$-$2005& 3686 & 2 & 0.90 & 1.185 \\    
18 & 24$-$11$-$2005& 3699 & 1 & 0.78 & 1.265 \\    
18 & 24$-$11$-$2005& 3699 & 2 & 0.90 & 1.443 \\    
19 & 06$-$12$-$2005& 3711 & 1 & 1.10 & 1.720 \\    
19 & 06$-$12$-$2005& 3711 & 2 & 1.09 & 1.445 \\    
20 & 24$-$05$-$2006& 3880 & 1 & 0.87 & 1.709 \\    
20 & 24$-$05$-$2006& 3880 & 2 & 0.90 & 1.443 \\    
21 & 16$-$06$-$2006& 3903 & 1 & 0.66 & 1.213 \\    
21 & 16$-$06$-$2006& 3903 & 2 & 0.51 & 1.155 \\    
22 & 20$-$06$-$2006& 3907 & 1 & 0.64 & 1.286 \\    
22 & 20$-$06$-$2006& 3907 & 2 & 0.58 & 1.189 \\    
23 & 27$-$06$-$2006& 3914 & 1 & 0.41 & 1.145 \\    
23 & 27$-$06$-$2006& 3914 & 2 & 0.50 & 1.133 \\    
24 & 27$-$07$-$2006& 3944 & 1 & 0.74 & 1.316 \\    
24 & 27$-$07$-$2006& 3944 & 2 & 0.76 & 1.204 \\    
25 & 03$-$08$-$2006& 3951 & 1 & 0.73 & 1.246 \\    
25 & 03$-$08$-$2006& 3951 & 2 & 0.65 & 1.169 \\    
26 & 13$-$10$-$2006& 4022 & 1 & 0.59 & 1.176 \\    
26 & 13$-$10$-$2006& 4022 & 2 & 0.52 & 1.300 \\    
27 & 28$-$10$-$2006& 4037 & 1 & 0.57 & 1.148 \\    
27 & 28$-$10$-$2006& 4037 & 2 & 0.53 & 1.138 \\    
28 & 10$-$11$-$2006& 4050 & 1 & 0.89 & 1.515 \\    
28 & 10$-$11$-$2006& 4050 & 2 & 0.88 & 1.323 \\    
29 & 27$-$11$-$2006& 4067 & 1 & 0.87 & 1.255 \\    
29 & 27$-$11$-$2006& 4067 & 2 & 0.92 & 1.391 \\    
30 & 19$-$12$-$2006& 4089 & 2 & 1.04 & 2.125 \\    
31 & 22$-$12$-$2006& 4092 & 1 & 0.80 & 2.018 \\    
31 & 23$-$12$-$2006& 4093 & 2 & 0.76 & 2.248 \\    
\hline  					        
\end{tabular}					        
\end{flushleft} 				        
\end{table}					        
						        
We used the G300V  grism in  combination with the GG375 order sorting
filter.  For our slit width, the spectral resolution was $\Delta\lambda=15$~\AA, as 
measured from the FWHM of night-sky emission lines, and
the resolving power was  $R=\lambda/\Delta
\lambda \simeq 400$ at the central wavelength $\lambda=5900$~\AA.  
The useful wavelength  range was 3900$~<\lambda<~$8200~\AA\ with a scale 
of $2.69$~\AA\  per pixel in the  spectral direction.  
%
This configuration
favors spectral coverage  rather than spectral resolution, allowing us to
follow the continuum  over a broad spectral  range, starting  with the
very  blue portion of the optical spectrum. 
Even so and in spite of $R=400$,   a
detailed profile of the BEL is still accessible.


We also  observed spectrophotometric standard stars
(GD~108,  HD~49798, LTT~377,  LTT~1020,  LTT~1788, and LTT~7987)
to remove the response  of  the telescope, CCD, and 
grism.  We did the relative calibration  between the epochs    
using the PSF stars
(see for more details Eigenbrod et al. \cite{cosmograil2}).

\begin{figure}[th!]
\begin{center}
\includegraphics[width=9cm]{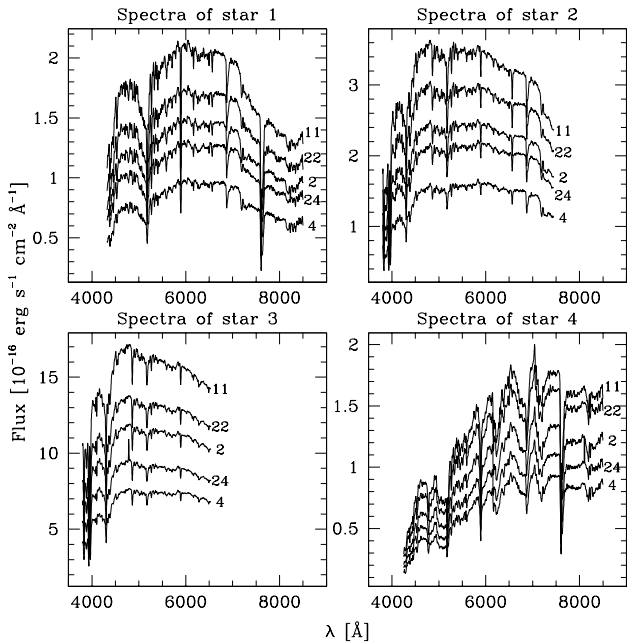}
\caption{Spectra of the four PSF stars. The spectra in each panel
correspond to different observing epochs, chosen to span the full length of the
monitoring. The IDs of the observing epochs, as given in Table~\ref{journal}, are
indicated.
The differences in flux are mainly due to the presence
of thin clouds. The purpose of using these stars as flux cross-calibrators
is precisely to eliminate these differences, both in intensity and shape.}
\label{psf_spectra}
\end{center}
\end{figure}

\begin{figure}[ht!]
\begin{center}
\includegraphics[width=9.0cm]{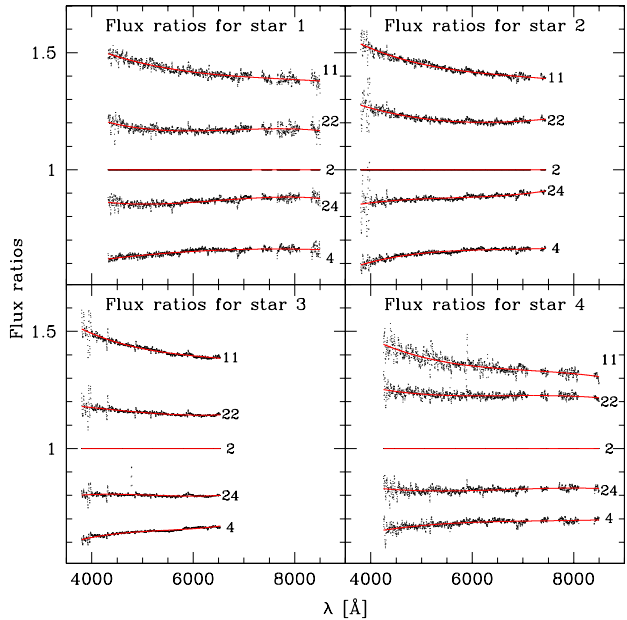}
\caption{Flux correction for different epochs with respect to the reference 
epoch \#~2 (14$-$11$-$2004).
Each panel corresponds to one of the 4 PSF stars visible in
Fig.~\ref{psf_spectra}. In each panel, the dotted line shows the ratio
of the spectrum of one PSF star taken  at  a given      epoch  and 
the spectrum of the same star taken at the reference
epoch \#~2. The curves are polynomials fitting the data.
Importantly, the correction derived  at a given  epoch  is  about the same
for the four stars. The mean  of these four curves is used
to correct the spectra  of the Einstein Cross images.   The small parts of
the   spectra with  strong atmospheric   absorption   are masked.  The
different spectral ranges are due to different clippings of the spectra
by the edges of the CCD.}
\label{psf_ratios}
\end{center}
\end{figure}

Finally, it is worth emphasizing that all our VLT data used 
in the present paper were obtained
in service mode, without which this project would have been impossible.

\section{Data analysis}

\subsection{Reduction}

The data reduction  followed the same procedure described in detail
in Eigenbrod et al.  (\cite{cosmograil3}).  We  carried out the standard
bias  subtraction and  flat   field correction of  the  spectra  using
IRAF\footnote{IRAF  is distributed by  the National  Optical Astronomy
Observatories,  which are operated  by the Association of Universities
for Research in Astronomy, Inc., under  cooperative agreement with the
National Science Foundation.}. We obtained the wavelength calibration from 
the spectrum of helium-argon lamps.  All spectra, for the object and for 
the PSF  stars were calibrated in two dimensions.

Only  one single exposure  was taken per  mask and per epoch.  For this
reason,  the usual  cosmic-ray rejection  scheme   applied to multiple
images  could not be applied. Instead, we used the L.~A. Cosmic algorithm
(van  Dokkum \cite{vandokkum01}), that  can handle single images.  We visually 
inspected the cosmic-ray corrected images to  check
that no  data  pixel was affected   by the process,  especially  in the
emission lines and in the data with the best seeing.

We removed the sky background in a different way in the spectra of the
PSF stars and in those of the gravitational  lens.  For the PSF stars,
which are small compared with the  slit length (19\arcsec), we used the
IRAF task  {\tt background}.  This task  fits a second order Chebyshev
polynomial in the spatial direction to the areas  of the spectrum that
are not illuminated by the object, and subtracts it from the data.  As
the  lensing  galaxy in \obj\ is  larger  than  the slit  length, this
procedure is not  applicable. Instead, we used the slits positioned on
empty sky regions of the FORS1 field of view,  and located next to the
gravitational lens. The sky was fitted to these slits and removed from 
the slit containing the images of \obj.

\subsection{Flux cross-calibration}

Once the cosmic rays and sky background were removed, we applied a flux
cross-calibration of    the  spectra as   described by   Eigenbrod  et
al.  (\cite{cosmograil2}), using the four  PSF  stars. The spectra of
these stars are shown in Fig.~\ref{psf_spectra} for five different observing 
epochs. Our observations show that these stars are non variable. 

We created a ratio spectrum for
each star, i.e.  we divided the spectrum of the star by the spectrum of
the same star for a chosen reference exposure.  We chose epoch \#~2 
(14$-$11$-$2004)
as our reference exposure because of the excellent weather conditions at
this  particular epoch for both seeing  and sky transparency.   The
computation of these flux ratios  was done for all  four stars in  each
exposure and we checked the compatibility of the response curves  derived with the four
different stars  are compatible  (see Fig.~\ref{psf_ratios}).  If not,
we rejected one or a maximum of two of the PSF stars. This can happen in
some  exceptional  cases, e.g. when the alignment  between  the star and the
slit is not optimal and generates a  color gradient in the spectrum of
the misaligned object. Aside from this instrumental effect, the observations 
show no trace of intrinsic variability of the PSF stars. 

After checking that the correction spectra obtained for the four stars
were  very  similar, we   computed their  mean,   which we  took as  the
correction to be applied to the gravitational lens. The high stability
of the corrections  across the field  demonstrates that  all residual
chromatic slit losses  due to  the  atmospheric refraction are   fully
corrected. This correction is eased by the  fact that: (1) the position
angle of the masks is the  same for the  quasar images and for the PSF
stars  (i.e. the PSF  clipping  is the same  for  the target and  the
reference stars); (2) we avoid observations at  large airmasses 
(i.e. never larger than 2.5); and (3)
the atmospheric refraction corrector on FORS1 is very efficient.

\begin{figure*}[t!]
\begin{center}
\includegraphics[width=13.cm]{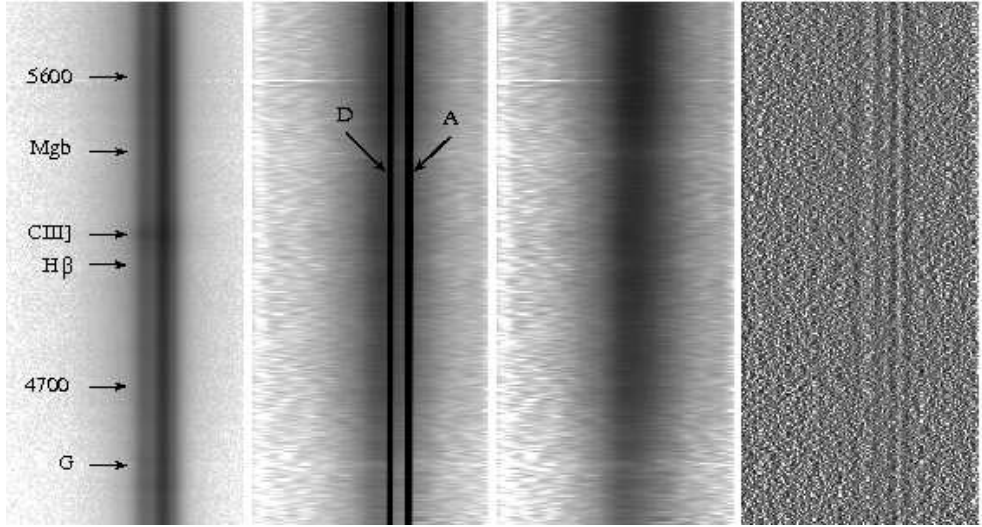}
\caption{
\emph{Left:} portion of the VLT 
2D-spectrum of quasar  images D and A, taken on epoch \#~25 (03$-$08$-$2006), on which
are indicated the main spectral features  of either the quasars or the
lens.  \emph{Center left:}  spatially   deconvolved spectrum.  
The two  quasar  images are very well
separated. \emph{Center right:} spectrum of the lensing galaxy alone.
\emph{Right:} residual map of the deconvolution after subtraction 
of the quasar and lens spectra. 
Note that the residuals are displayed with much narrower cuts than those used in the other panels. 
The darkest and brightest pixels correspond to $-3\sigma$ and $+3\sigma$ respectively.
No significant residuals of the spectral features are visible.}
\label{2D_spectra}
\end{center}
\end{figure*}

\begin{figure*}[t!]
\begin{center}
\includegraphics[width=15.4cm, height=10cm]{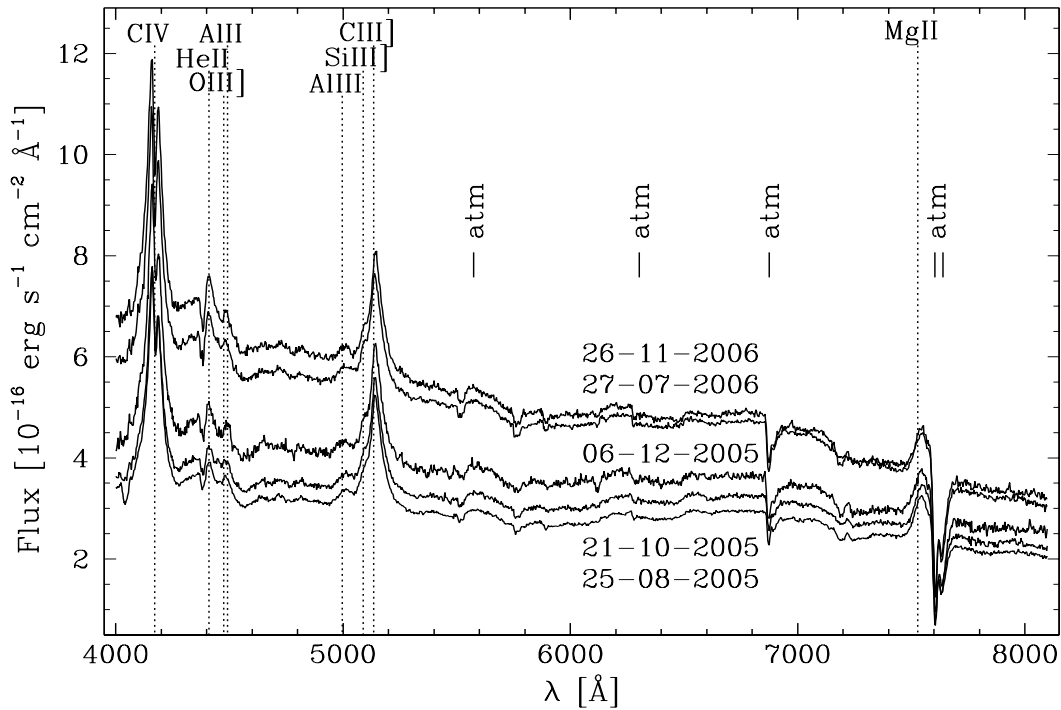}
\caption{
Deconvolved and extracted spectra of quasar image A for five observing epochs. 
Chromatic variations in the spectra
are conspicuous with the blue part of the spectra being more magnified than 
the red part.}
\label{A_spectra}
\end{center}
\end{figure*}

\begin{figure}[t!]
\begin{center}
\includegraphics[width=9.0cm]{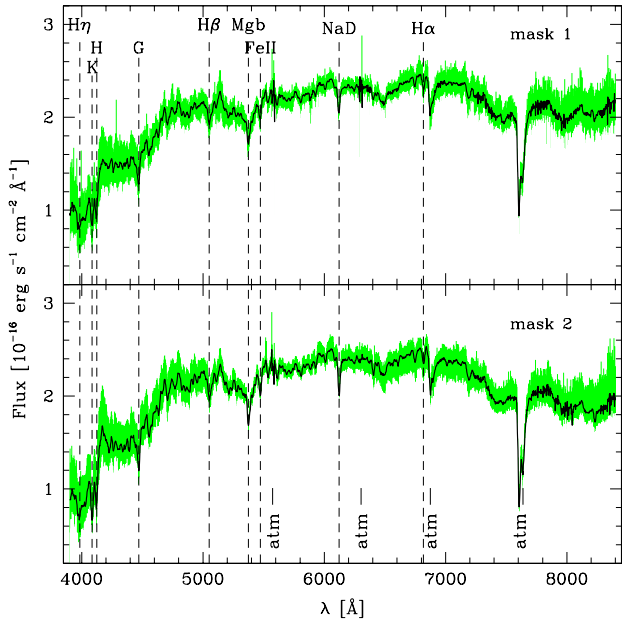}
\caption{Deconvolved and extracted 1D-spectra of the lensing galaxy.
The two panels correspond to the two MOS masks. The shaded areas are the
envelopes  containing all the  spectra  of the  lens obtained with the
corresponding mask.  The thick black lines are the means. Note the small
scatter between the two spectra.}
\label{lens_spectra}
\end{center}
\end{figure}

\begin{figure}[t!]
\begin{center}
\includegraphics[width=9.0cm]{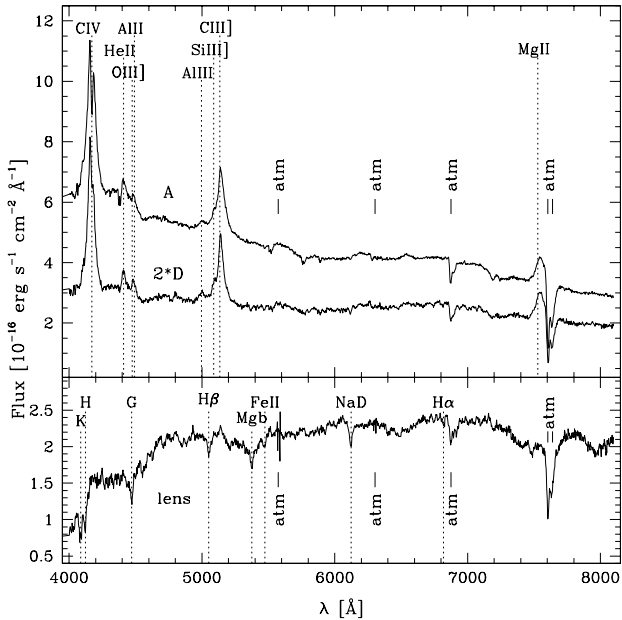}
\caption{Example of a spectral decomposition. The top panel shows 
the two extracted spectra for the images A and D for the observations taken on
epoch \#~25 (03$-$08$-$2006) with mask 1.   The  extracted spectrum  of the lensing
galaxy, in the   bottom panel, shows  no  trace of contamination by  the
quasar   BELs.   For   clarity, we have
multiplied the flux of image D by a factor of two.}
\label{mask_spectra}
\end{center}
\end{figure}

\subsection{Deconvolution}
 
The lensing galaxy in \obj\ is bright. Its central parts have
a surface brightness of approximately 18~mag/arcsec$^2$ in the R band, 
which is comparable to the quasar images. Hence studying microlensing
variations of the  quasar images requires  very accurate deblending.
 
In order  to  carry  out this  challenging task,  we  used the spectral
version  of the MCS deconvolution   algorithm (Magain et al.
\cite{magain98}, Courbin et al.   \cite{courbin}), which uses
the spatial information  contained in the  spectra of the PSF stars.  The
algorithm sharpens the spectra  in  the  spatial direction, and   also
decomposes them into a ``point-source channel'' containing the spectra
of the two  quasar images, and an  ``extended channel'' containing the
spectrum of  everything in the  image that is  not a point source, in
this case, the spectrum  of the lensing galaxy.  In Fig.~\ref{2D_spectra},
we illustrate      the   process   and     the   different      outputs.
In Fig.~\ref{lens_spectra}, we show how similar   the spectra of   the
lensing galaxy are, extracted  either from two different masks or from
data taken at different epochs, hence illustrating the robustness of the 
deconvolution technique. In Fig.~\ref{mask_spectra}, we give an example of
decomposition of the data into the quasar and  lens spectra after 
integrating along the spatial direction. The lensing galaxy spectrum
shows no trace of the residual quasar BELs. Even when the
contrast between the quasar and the galaxy is particularly large, the decomposition
is accurate. For example, the CaII H+K doublet in the lens spectrum is well
visible, in spite of the presence of the strong quasar \ion{C}{IV} emission in the
same wavelength range.

\begin{figure}[t!]
\begin{center}
\includegraphics[width=9.0cm]{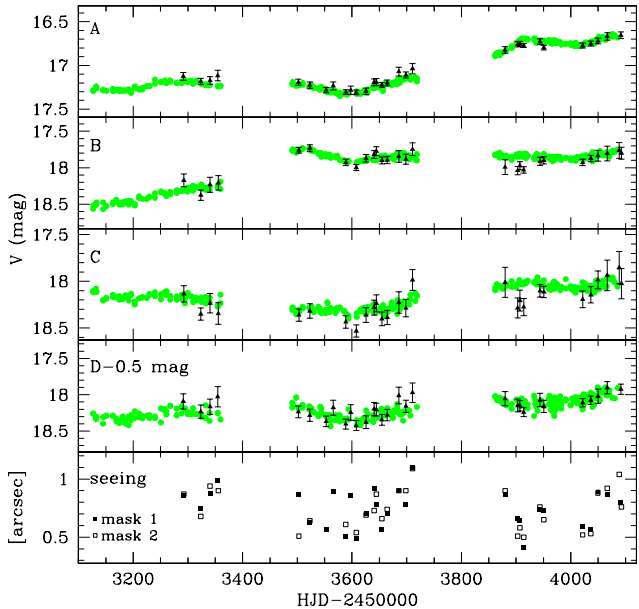}
\caption{OGLE-III light curves (Udalski et al. \cite{udalski}) of all four quasar images 
from  April 2004  to  December 2006   (dots),  compared with the
photometry derived by   integrating our VLT  spectra through  the OGLE
V-band (dark triangles).  The $1-$sigma  error bars correspond to the
photon noise in the  spectrum.  We shift   the OGLE-III light curve  of
image  D   by $-0.5$~mag with  respect to   the published  values.
The bottom panel displays the seeing values for each observations.}
\label{ogle}
\end{center}
\end{figure}

\subsection{Cross-check with the OGLE-III light curves}

After  reduction and  spatial  deconvolution, we  obtained the  extracted
spectra of quasar images A and D on 30 different epochs, and of B and C
on 28  different epochs. Several extracted spectra of image A are shown in 
Fig.\ref{A_spectra}.
As a sanity  check, we compared our results with
the OGLE-III photometric monitoring of QSO~2237$+$0305 (Udalski et al.
\cite{udalski}).  We integrated our quasar spectra in the corresponding
$V$-band to estimate, from the spectra, the photometric 
light curves as if they were obtained from imaging. In Fig.~\ref{ogle}, we compare
our magnitude estimates with   the  actual OGLE-III measurements.   The  overall
agreement is very good for images A, B, and C.  For image D, we have
to  shift the  OGLE-III light curve by  $-0.5$~mag with  respect to the
published   values. 
Interestingly, this shift is not needed when we compare our
results with the previous OGLE data from the provisional calibration presented 
in the years 2004$-$2006. The previous OGLE data also agreed with the photometry 
of Koptelova et al. (\cite{koptelova}). 
This changed when Udalski et al. (\cite{udalski}) reviewed their 
calibration and gave image D a larger magnitude of approximately $0.5$~mag.
They stated that the steep rise of brightness of image D
at the end of the 2000 OGLE-II season leaded to an overestimate of the 
extrapolated magnitude for the beginning of the 2001 OGLE-III season. 
But this is now discrepant with the photometry of 
Koptelova et al. (\cite{koptelova}). 
We think that the new extrapolation of the light curve of image D from the end of season
2000 to the beginning of season 2001 might be uncertain, leading to the observed
shift between our data and the OGLE-III light curve of image D. 
However, aside from this  shift,  the agreement between the
OGLE photometry and our integrated VLT spectra is also very  good for image D.

\begin{figure*}[p!]
\begin{center}
\includegraphics[width=17cm, height=11.05cm]{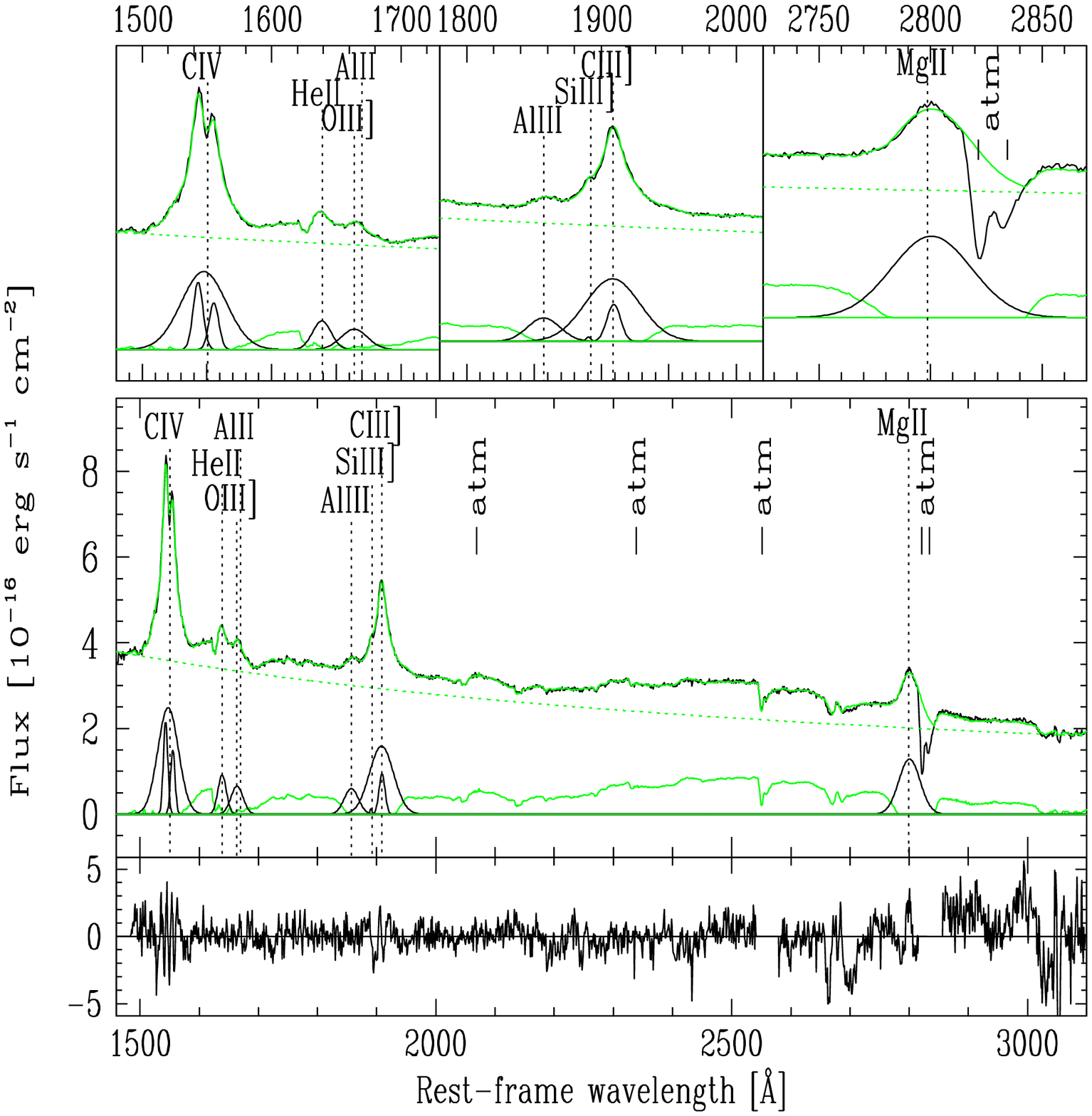}
\caption{Multi-component decomposition of the spectrum of the brightest quasar 
image,  A, taken on epoch \#~2 (14$-$11$-$2004).  The  upper  panels show  the
detailed spectral decomposition of the BELs, while the
middle panel displays the entire spectrum.  The continuum is indicated
as a   dotted  curve. The Gaussian   lines  and  iron pseudo-continuum
templates are shown below the  spectrum.  The bottom panel is the residual
for each pixel normalized by the photon noise per pixel (i.e. the 
y-axis is the residual flux in units of $\sigma$).}
\label{mcd}
\vskip 20pt
\includegraphics[width=8.9cm]{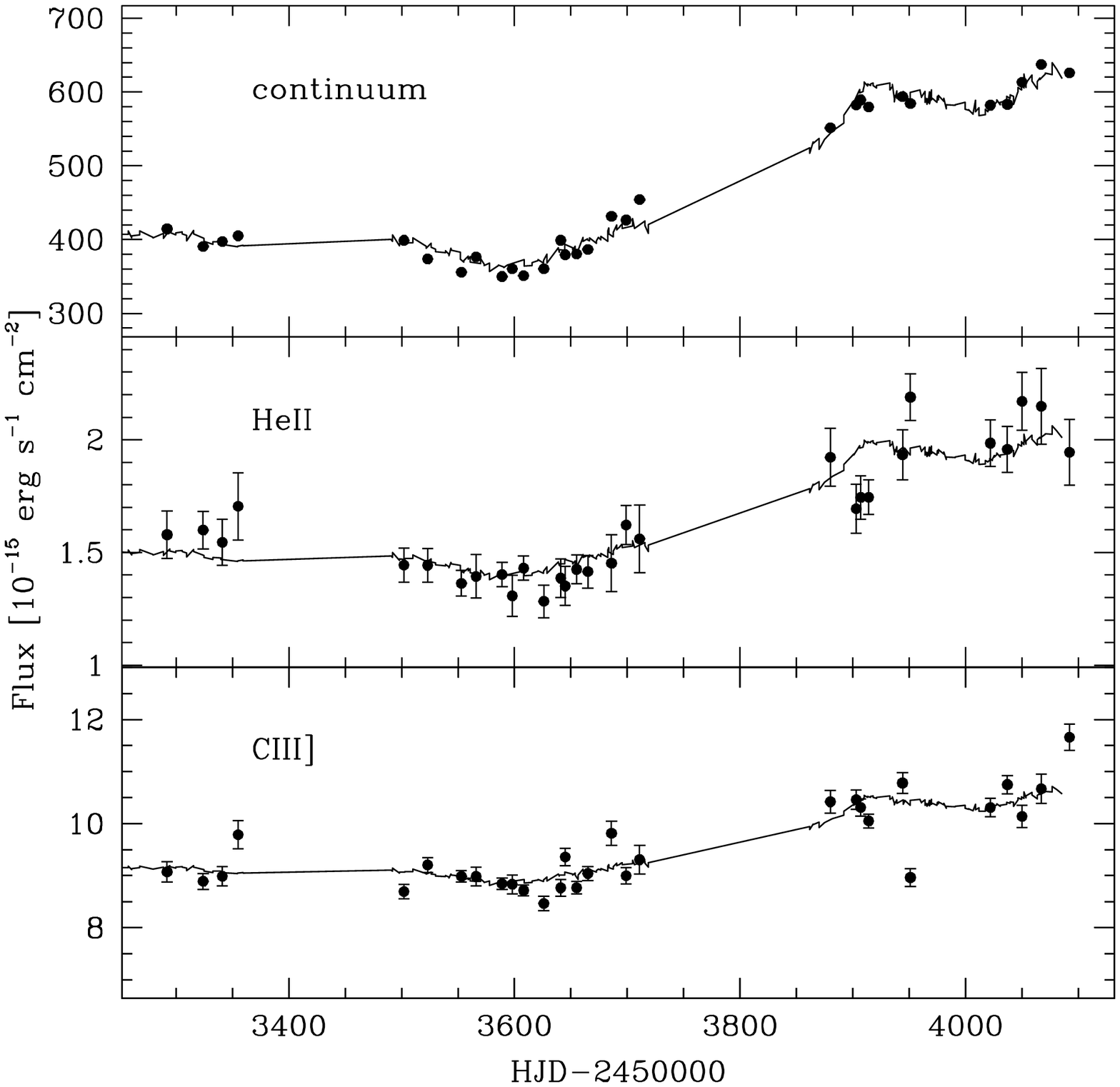}
\leavevmode
\includegraphics[width=8.9cm]{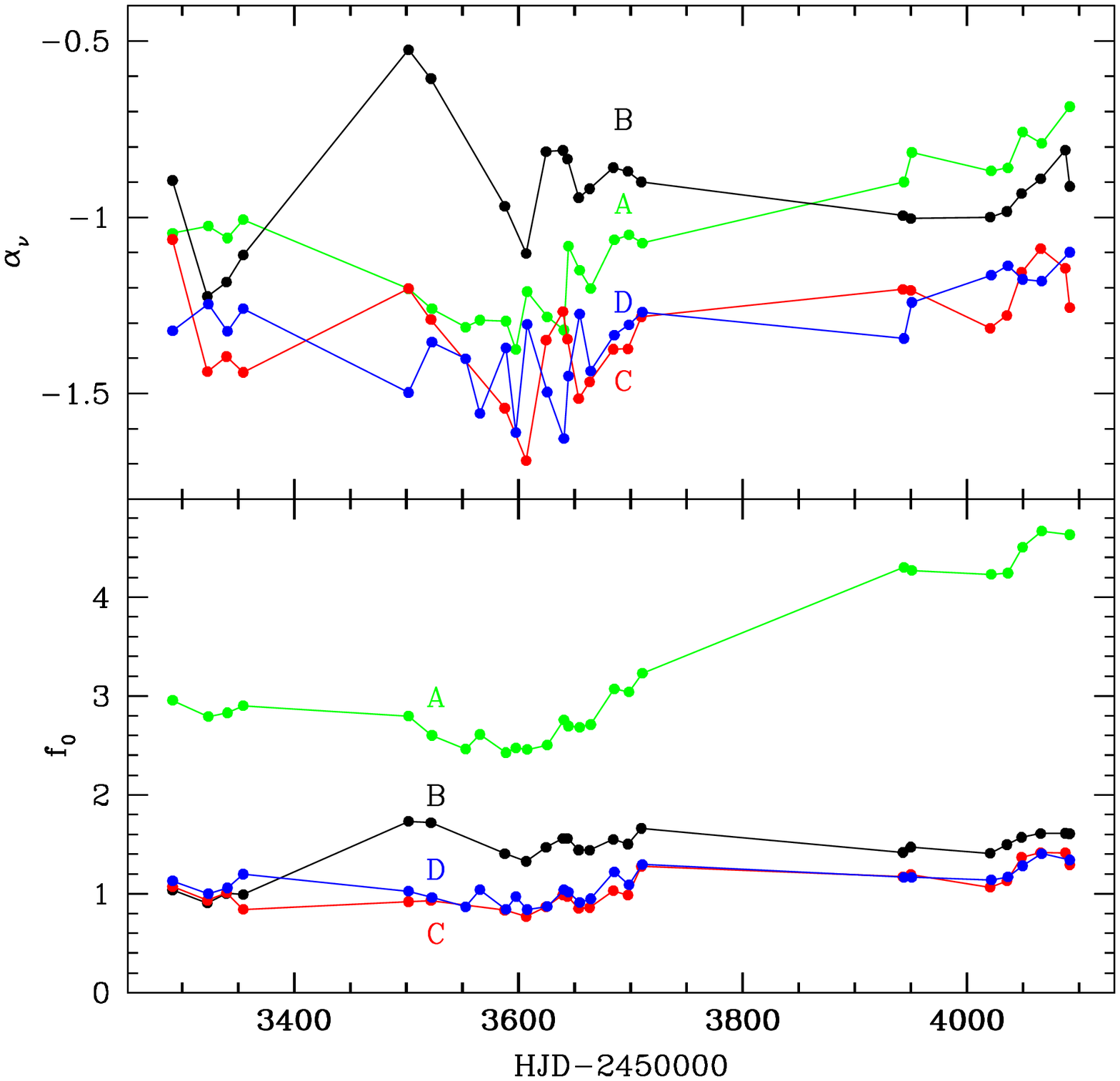}
\caption{\emph{Left:} Examples of light curves for the
quasar image~A. The  integrated flux 
for the   continuum, the \ion{He}{II}, and   \ion{C}{III]}  BELs
are given  from top to bottom.  
The continuum is integrated over the entire available wavelength range.
In each panel, we fit a scaled version (solid line) of the
OGLE-III light curve  (Udalski et al. \cite{udalski}).  
This  nicely  illustrates that  the BELs  vary
simultaneously and  proportionally  to the  continuum.   \emph{Right:}
variability of the  best-fit parameters $\alpha_{\nu}$ and $f_0$ of the continuum
(see Section~\ref{multicomp}).  Note  how  the  changes  in slope  of   the
continuum ($\alpha_{\nu}$) follow those of the mean intensity ($f_0$).}
\label{lc}
\end{center}
\end{figure*}

\section{Multi-component decomposition}

Different emission  features are known   to be produced  in regions of
different  characteristic  sizes. As microlensing 
magnification varies on short spatial scales, sources of different sizes 
are magnified by differing amounts (e.g. Wambsganss et al. \cite{wambsganss}). 
Emission features from smaller regions of the source  are more highly variable due to 
microlensing than features emitted in more extended regions.  
In order to study the variation
of each spectral  feature independently,   we  need to decompose   the
spectra into their individual components.


\subsection{Method}
\label{multicomp}

In our analysis of the 1-D spectra of the four quasar images, we
follow the multi-component decomposition (MCD) approach (Wills et al.
\cite{wills}, Dietrich et al. \cite{dietrich}) implemented in Sluse
et al. (\cite{sluse}).  This method is applied to the rest-frame
spectra, assuming they are the superposition of (1) a power law
continuum, (2) a pseudo-continuum due to the merging of \ion{Fe}{II}
and \ion{Fe}{III} emission blends, and (3) an emission spectrum due to the
other individual BELs.  We consider the following emission lines :
\ion{C}{IV}~$\lambda 1549$, \ion{He}{II}~$\lambda 1640$,
\ion{O}{III]}~$\lambda 1664$, \ion{Al}{II}~$\lambda 1671$,
\ion{Al}{III}~$\lambda 1857$, \ion{Si}{III]}~$\lambda 1892$,
\ion{C}{III]}~$\lambda 1909$, and \ion{Mg}{II}~$\lambda 2798$.  All
these features are fitted simultaneously to the data using a standard
least-square minimization with a Levenberg-Marquardt based algorithm
adapted from the Numerical Recipes (Press et al. \cite{press}).

In the  first step, we identify  the underlying  nonstellar power-law
continuum  from spectral  windows that  are free  (or almost free)  of
contributions  from  the     other  components,   namely   the    iron
pseudo-continuum     and   the      BELs. We     use    the    windows
$1680\le\lambda\le1710$~\AA\  and   $3020\le\lambda\le3080$~\AA. After
visual    inspection of the    iron     templates by Vestergaard    et
al.    (\cite{vestergaard}), we do   not   expect significant  iron
emission in these windows.

We characterize the spectral continuum (measured in the restframe) with a power law
$f_{\nu}     \propto \nu^{\alpha_\nu}$, which translates in wavelength to
$f_{\lambda} \propto \lambda^{\alpha_\lambda}$ with the relation
$\alpha_\nu=-(2+\alpha_\lambda)$, i.e.
$$f_{\lambda} = f_0              
\left(\frac{\lambda}{\lambda_0}\right)^{\alpha_{\lambda}}  =  f_0    
\left(\frac{\lambda}{\lambda_0}\right)^{-(2+\alpha_{\nu})}   
$$
where
$\lambda_0=2000$~\AA\ and  where $\alpha_{\nu}$  is the commonly  used
canonical power index.

Next, we fit  the BELs with Gaussian  profiles.  We consider a sum of
three  profiles to   fit  the absorption  feature in  the  \ion{C}{IV}
emission  line. Two profiles are  used for the \ion{C}{III]} line and
one single profile is used to fit simultaneously the \ion{O}{III]} and
\ion{Al}{II}   lines.   All other  BELs   are  fitted with one  single
profile.   We  then subtract   the  BELs and   the continuum from  the
spectra. We consider the  residuals as coming  from the emission blends
of  \ion{Fe}{II} and \ion{Fe}{III}.  Hence the averaged and normalized
residuals over all  epochs define our first iron pseudo-continuum
template.   We can   then  proceed   iteratively by  including    this
pseudo-continuum  iron  template  in the  fitting procedure  and
rerun it. This  gives  a better fitting  of   the emission  lines  and
defines  an   improved  iron pseudo-continuum template.  After
five  of  these  iterations, typically the fitting  does  not  change
significantly anymore.  Fig.~\ref{mcd}   shows an example  of the
fitting decomposition described above.

\begin{figure}[t!]
\begin{center}
\includegraphics[width=9cm]{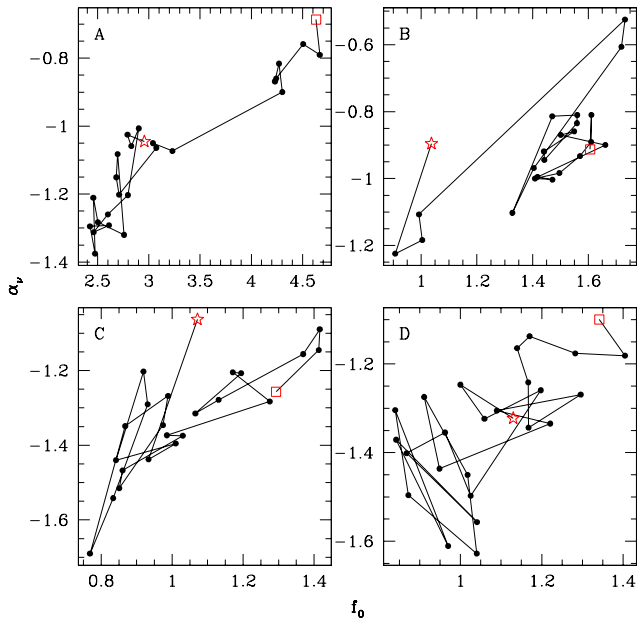}
\caption{Correlation between the intensity $f_0$
and slope $\alpha_{\nu}$ of the continuum  spectra for all four quasar
images. The points are connected chronologically. 
The first observation epoch is marked by a star and the last one 
by a square. The correlation  is  obvious in   images A  and B
spanning a broad range of spectral slopes. In these two images, an
increase in intensity is accompanied by an increase in steepness, i.e.
when a quasar image gets brighter, it also gets bluer.}
\label{alpha_f0}
\end{center}
\end{figure}

\subsection{Results}

The light curves  for the continuum and  for the emission lines can be
constructed from the above multi-component  decomposition.  We show in
Fig.~\ref{lc}  an example of variation  in the brightest quasar image,
A, for the continuum and for two BELs. The error bars
give the photon   noise, integrated over the  corresponding  wavelength
range.  In the right panel  of Fig.~\ref{lc}, we show the variability
of  the continuum in  intensity, $f_0$,  and in slope, $\alpha_{\nu}$,
for  the 4 quasar  images.
It  is immediately clear that the continuum
variations  with the  largest amplitude are   observed in image A,  
between HJD~$=$~3600 and 3900~days, and in  image  B between 
HJD~$=$~3300 and 3500~days. These
variations are accompanied by an  increase in  steepness, i.e. when  a
quasar image  gets brighter, it also  gets bluer. 
This is particularly obvious in Fig.~\ref{alpha_f0}, where $f_0$ and 
$\alpha_{\nu}$ are strongly correlated for images A and B, indicating
that significant microlensing events occured in these two images.


Inaccurate alignment of the   quasar images in   the slit of  the
spectrograph   is  a possible instrumental    effect   that can  mimic
microlensing changes in the spectral slope of the quasar images.  Indeed, small
clipping of  one of the  quasar images would  lead  to a stronger flux
loss at the bluer wavelengths, hence producing a color gradient in the
spectrum and a  decrease in  the  measured value $\alpha_{\nu}$.   We
have  checked all the  ``through-slit'' images  taken before  each
spectrum.  Not   only do these
images  show  that the alignment is  correct,  but  it   is also very  easily
reproducible from one  epoch to another,  even when the FORS1 has been
dismounted from and remounted on the telescope.

We have also checked that our fitting procedure does not introduce any
spurious correlation between $\alpha_{\nu}$ and $f_0$. We check this 
by using simulated spectra.  In order to do that we
take a reference spectrum for each quasar image and subtract its
continuum.  We then take random pairs of ($\alpha_{\nu}$, $f_0$)
parameters so that the $\alpha_{\nu}$ vs. $f_0$ plane is well sampled.
We chose 400 such pairs and create the corresponding continuum to be
added to the reference spectrum. The decomposition procedure is then
run on the 400 spectra.  We find no correlation at all between the
measured $\alpha_{\nu}$ and $f_0$. In addition, the parameters used to
build the simulated spectra are almost perfectly recovered by the
decomposition procedure.

We conclude that genuine chromatic  variations are present in the
continuum of  all   images of  \obj.    The effect  is   most
pronounced  in image  A during  the last  observing season, and in
image  B at  the   beginning of our  monitoring.   We  show in the
following that these   observed   variations are, in  addition,   well
compatible with the OGLE-III single-band photometric observations.


\section{Microlensing variability in the OGLE-III photometry}
\label{micro_ogle}

The photometric variations  in most gravitationally-lensed quasars are
dominated by the intrinsic variations of the quasar,
typically of the order of $0.5-1.5$~mag, hence making them
useful to measure the time delays between the quasar images.
Microlensing variations are usually smaller, in the range  $0.05-0.1$~mag
(e.g. the lensed quasars 
B~1600$+$434 by Burud et al. \cite{burud00};
RX~J0911$+$0551 by Hjorth et al. \cite{hjorth}; 
SBS~1520$+$530 by Burud et al. \cite{burud02}; 
FBQ~0951$+$2635 by Jakobsson et al. \cite{jakobsson}).

The Einstein  Cross is different from
this general behavior in two ways: (1) the  time delays 
between each pair of images are expected to be 
of the order of one day, hardly measurable,
and (2) the microlensing variations dominate the 
light curves.  
For these two reasons, microlensing can be fairly well isolated 
in each quasar image, because   it  acts
differently  on the  four sightlines.

\begin{figure}[t!]
\begin{center}
\includegraphics[width=9cm]{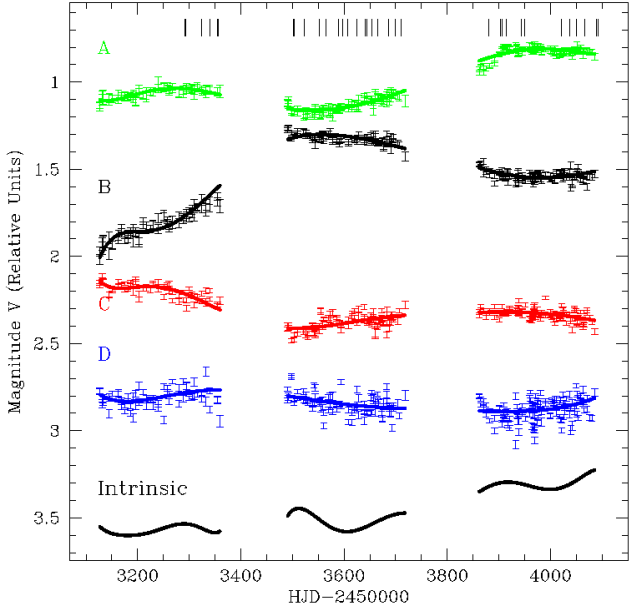}
\caption{Decomposition of the OGLE-III photometric light curves  
(Udalski et al. \cite{udalski}) of the quasar 
images,  into  intrinsic  quasar  variations  and  microlensing-induced
variations (see Section~\ref{micro_ogle}).  The intrinsic variations 
are shown at the bottom
of  the figure  as a  continuous  line,  while the  pure   microlensing
variations are  the data  points.  The curves are  shifted arbitrarily
along the y-axis for clarity. The tickmarks at the top show the epochs 
of our  observations. }
\label{micro}
\end{center}
\end{figure}

To separate the intrinsic flux  variations of the quasar from
the  microlensing  ones, we perform  a  polynomial fit to the OGLE-III
light curves (Udalski et al. \cite{udalski}) of   Fig.~\ref{ogle}.  
This simple  and fully  analytical
method has been developed by Kochanek et al.  (\cite{kochanek06}),
and  is also described by  Vuissoz  et al.  (\cite{vuissoz}).  In  the
present  application, the variations of  each quasar image are modeled
as a sum of two Legendre polynomials:  one polynomial is common to all
four   quasar images and represents the   intrinsic  variations of the
source  while a  second  polynomial, different  for  each  quasar image,
represents the additional microlensing variations. In doing so, we 
rescale the OGLE-III error bars of each image by a factor equal
to the flux ratio between each image and image C. This 
rescaling suppresses the potential problem existing if the 
fitting procedure considers the variation
of image A (with the highest signal-to-noise) as the intrinsic
variation of the quasar. The chosen order of the polynomial is 7 for the
intrinsic variation, and 10 for the microlensing variation. Higher order
polynomials do not significantly improve the fit. 
The results  are  displayed in  Fig.~\ref{micro},  where the intrinsic
variation of the  source recovered  by  the simultaneous fit  is shown
together   with  the pure  microlensing   variations.

We check  the
efficiency of our method by generating artificial light curves and then
using     the above  polynomial fit   to    recover the intrinsic  and
microlensing light curves.  These artificial light curves are generated
in   the   same   way  as     described   in Eigenbrod    et   al.
(\cite{cosmograil1}), and are composed of an  intrinsic light curve to
which we add microlensing fluctuations. Both are created in a random walk manner
(i.e.  not from polynomials).  They are  constructed to match the
variability properties of  the actual light curves,  i.e.  their
timescale and    amplitude of variation (for further details see Eigenbrod    et   al.
\cite{cosmograil1}).    We recover  the simulated
intrinsic light curve of the  quasar with a typical  error of less than
0.1~mag. The variations of more than 0.4~mag, shown in Fig.~\ref{micro},
both for microlensing and quasar  variations, are well above the error
estimated from  the simulated  light curves. In  our simulations,  we
adopt the same photometric error bars as in  the light curves of all
quasar images, i.e. the re-scaling of  the error bars described above
in the real data is taken into account. If, on the contrary, we adopt 
error bars that follow the photon noise, the fitting procedure considers
the highest signal-to-noise light curve as the intrinsic quasar
light curve.

The  light curve  most   affected  by microlensing   is  that of
ima\-ge~B, with a peak-to-peak  amplitude of more  than 0.7~mag over 3
years.  The  other quasar images  show microlensing-induced variations
of up to 0.4~mag, with quasar image A  having a sharp event
during the  last observing season.   The intrinsic quasar light curve
displays a variation of about 0.4~mag.

The polynomial decomposition of the  light curves are  compatible
with the spectroscopic results.  Quasar images A and  B, which have the
largest microlensing contribution in Fig.~\ref{micro}, respectively at
HJD~$\sim$~3500 days  and HJD~$\sim$~3900 days, also  have a sharp rise in
$\alpha_{\nu}$ at the same epochs.

\section{Microlensing variability in the spectra}

Chromatic variations  of the continuum of  images A and B of \obj\
are clearly seen in  our data. In addition, differential magnification
of  the continuum with respect to  the BELs is   also seen in all four
quasar  images.   Such effects have  already  been observed  by  Lewis  et al.
(\cite{lewis98})  and Wayth et     al.  (\cite{wayth}), but  only  for
data over two epochs.  Our VLT  spectra allow us to  follow the variations over
two  full years, provided the intrinsic  variations of  the quasar are
removed.

\subsection{Continuum and BELs relative magnifications}

Since  the time delays  in  \obj\  are  negligible,  taking  the ratio
between   the above quantities in pairs   of quasar images cancels the
intrinsic   variations. Let $F(t)$  be  the intrinsic source flux, and 
$M_i$, $\mu_i$ be   the macro and microlensing-magnifications of 
quasar image $i$, respectively. 
The observed flux ratio between images $i$ and $j$ at time $t$ is then :

\begin{equation}
R_{ij}(t) = \frac{\mu_i(t) \ M_i \ F(t)}{\mu_j(t) \ M_j \ F(t)} 
~ e^{-\left(\tau_i -\tau_j\right)}
= \frac{\mu_i(t) \ M_i }{\mu_j(t) \ M_j }
~ e^{-\left(\tau_i -\tau_j\right)} ~\textrm{ .}
\end{equation}

The extinction $e^{-\tau_i}$ remains constant in time and is relatively 
similar in all four quasar images as we show in Sec.~\ref{extinction}. 
Hence, it is not expected to strongly
affect our results and we will neglect it in the following.
The macromagnifications $M_i$ are best estimated in the mid-IR and radio
domain (Falco et al. \cite{falco}, Agol et al. \cite{agol00}). At these wavelengths,
the source  size is much larger than  the typical spatial scale in the
microcaustics network,     hence  leaving it   fairly   unaffected  by
microlensing (i.e. $\mu_i =1$). By multiplying $R_{ij}$ by $M_j/M_i$,
using the mid-IR observations, we find the pure-microlensing magnification
ratios:

\begin{equation}
r_{ij}(t) = \frac{\mu_i(t)}{\mu_j(t)}	~\textrm{ .}
\label{mumu_eq}
\end{equation}

\begin{figure*}[ht!]
\begin{center}
\begin{tabular}{ccc}
\includegraphics[height=3.6cm,width=5.1cm]{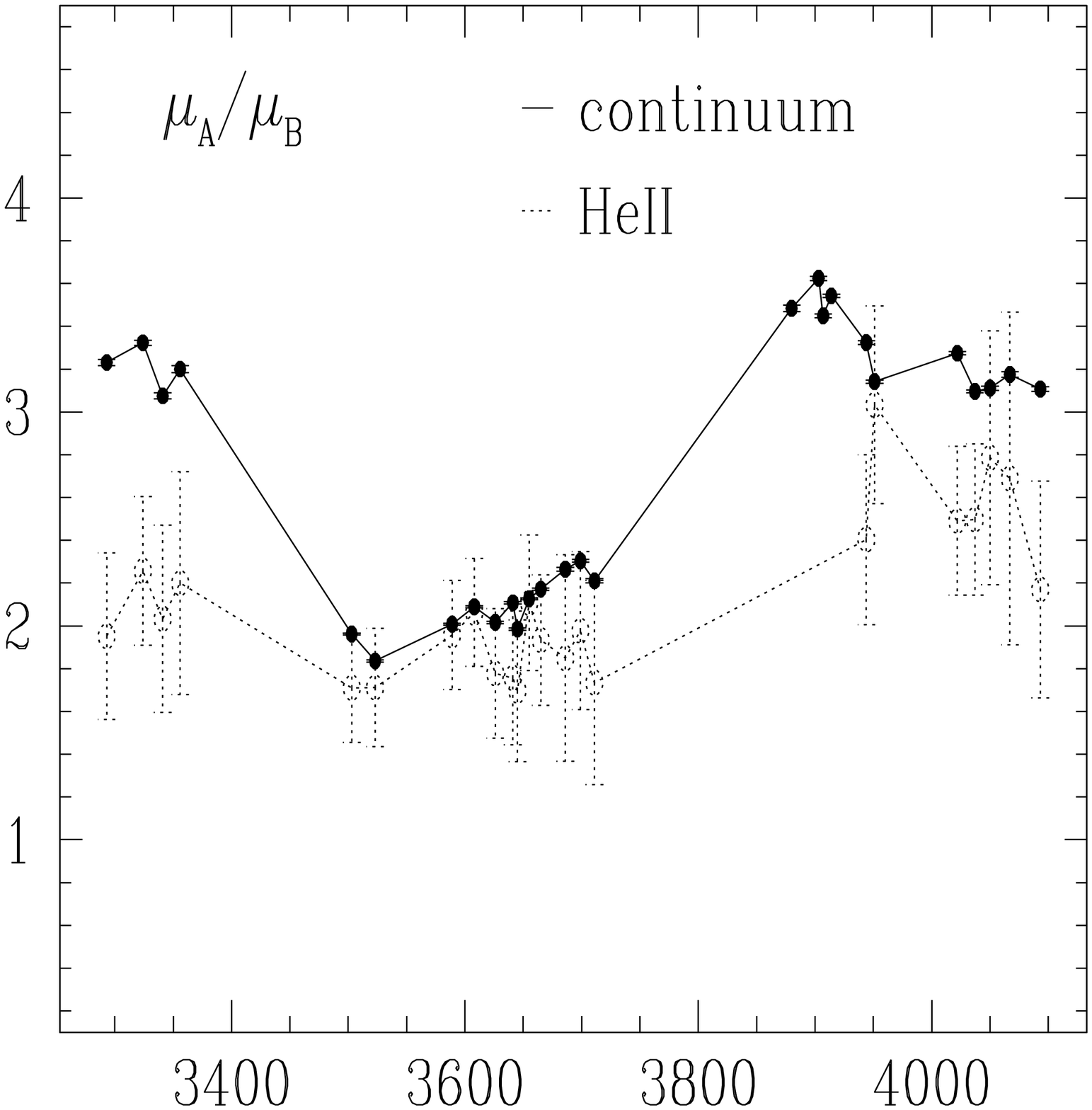}&
\includegraphics[height=3.6cm,width=5.1cm]{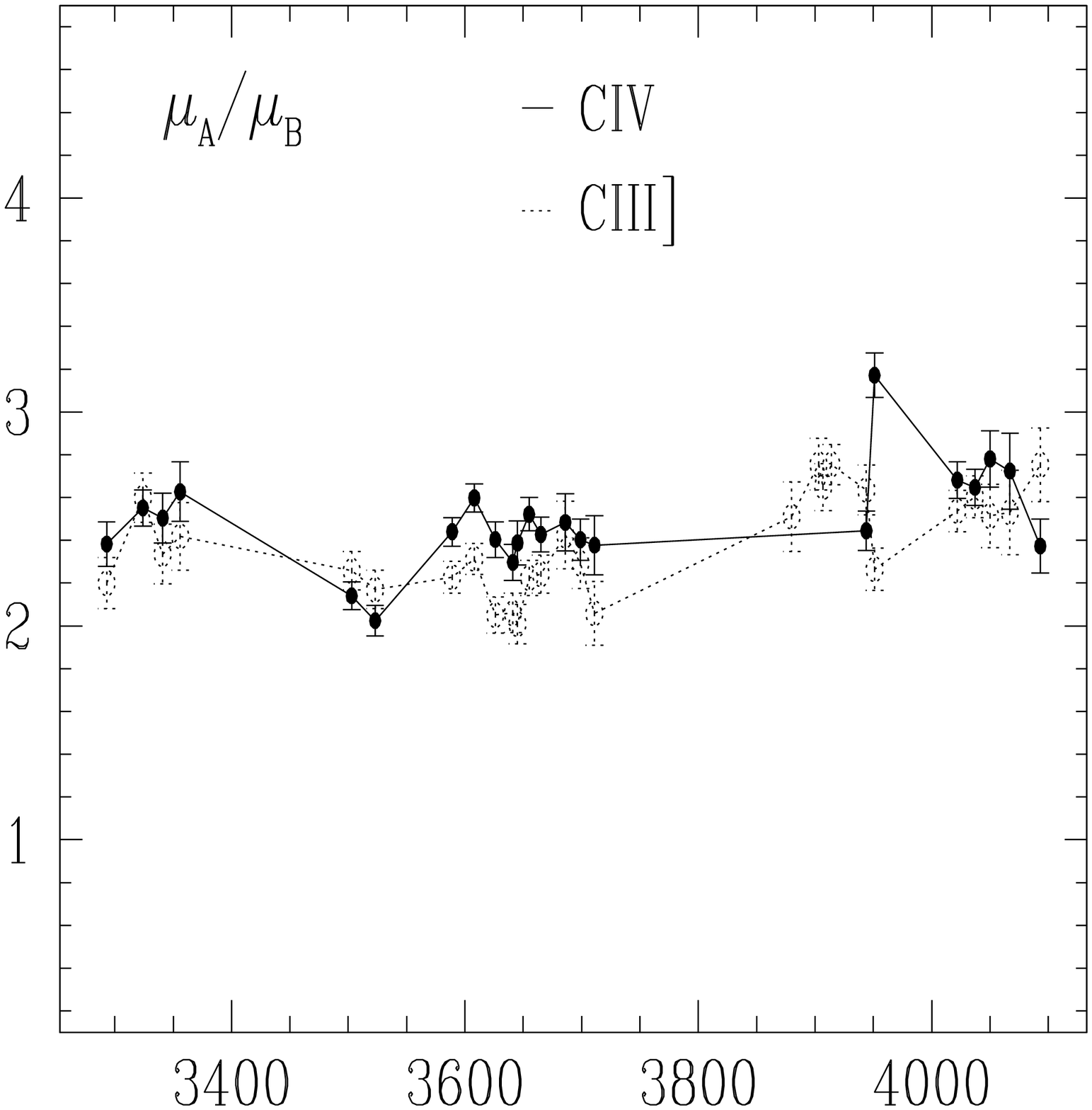}&
\includegraphics[height=3.6cm,width=5.1cm]{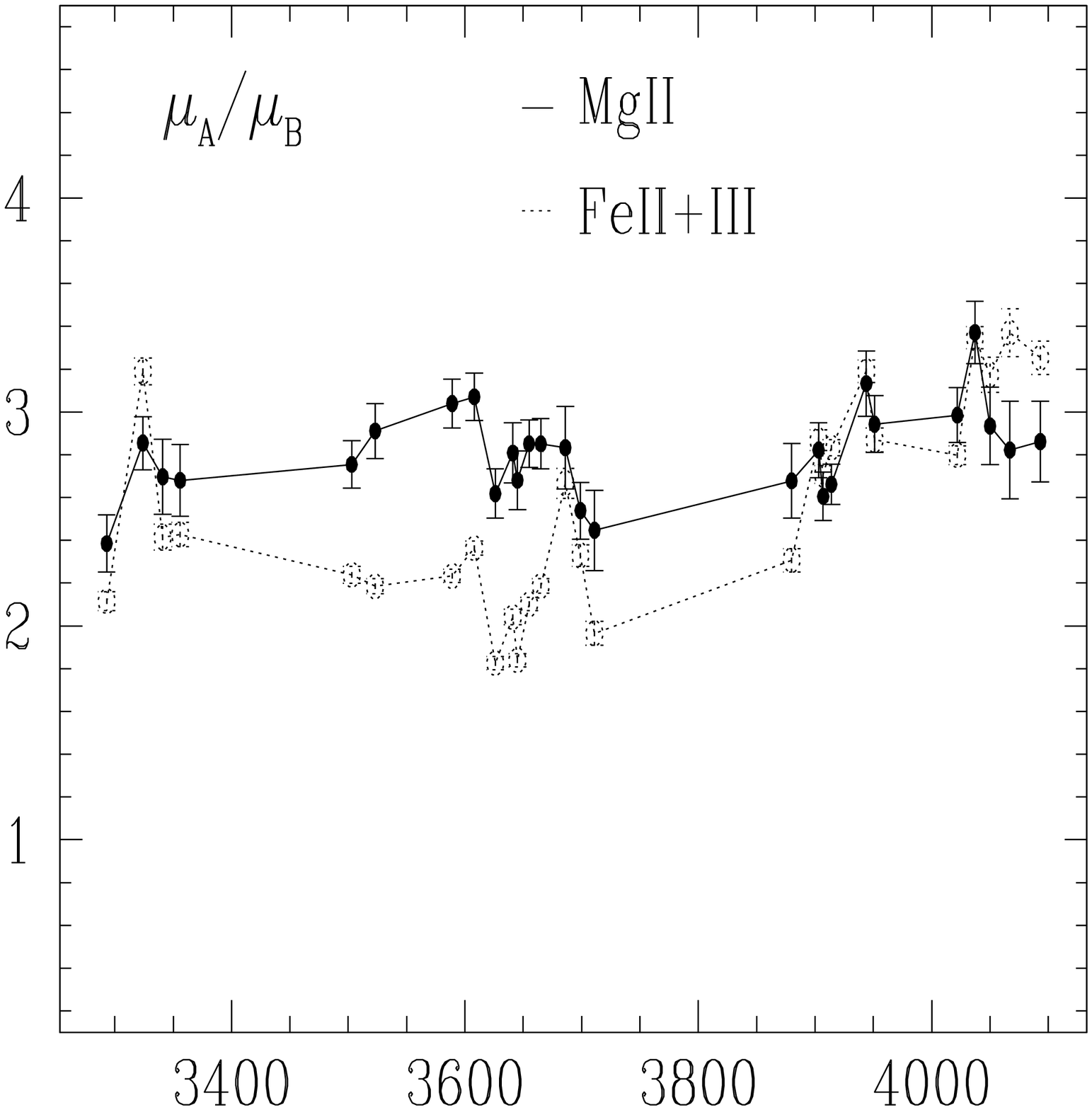}\\
\includegraphics[height=3.6cm,width=5.1cm]{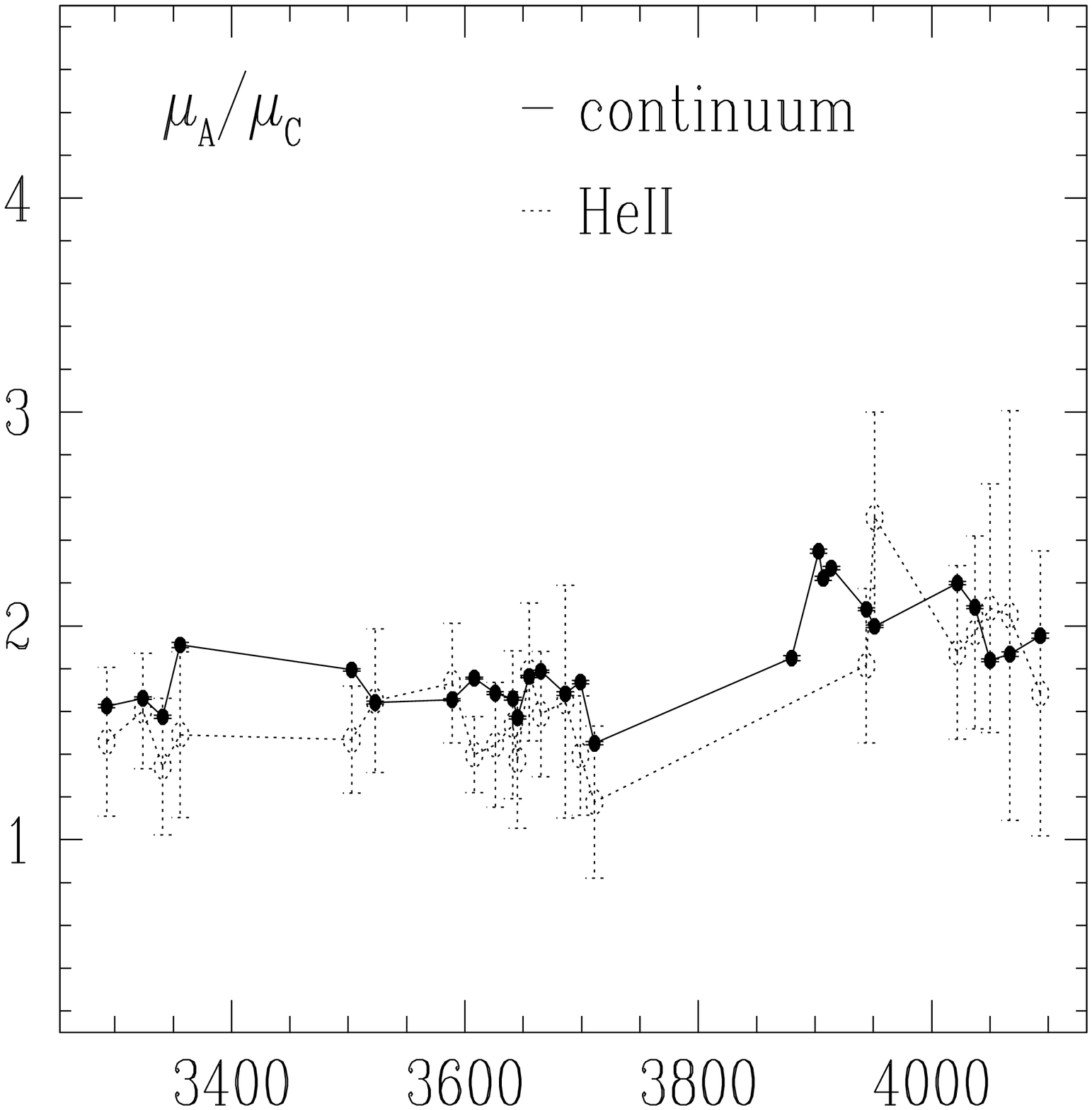}& 
\includegraphics[height=3.6cm,width=5.1cm]{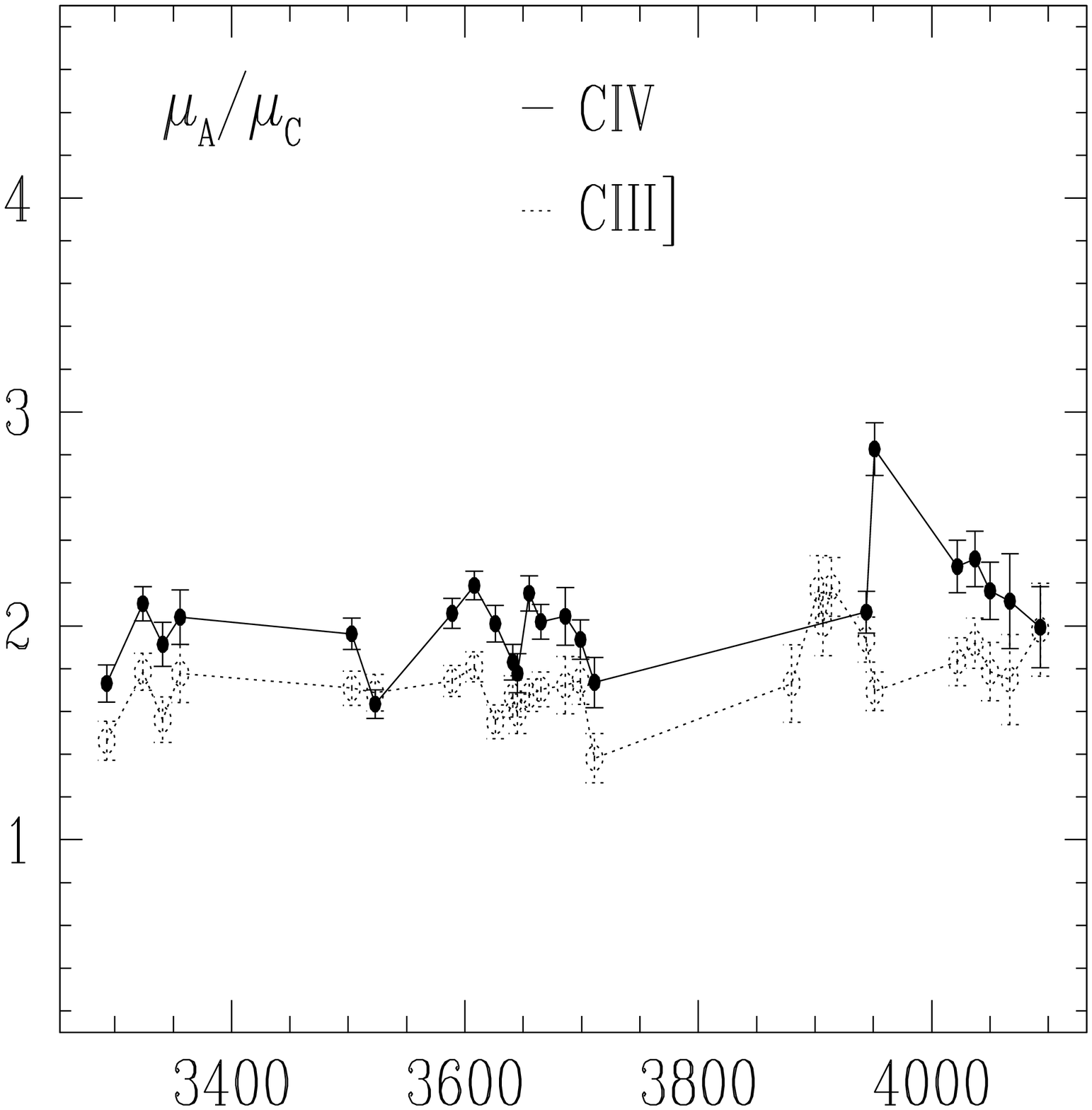}& 
\includegraphics[height=3.6cm,width=5.1cm]{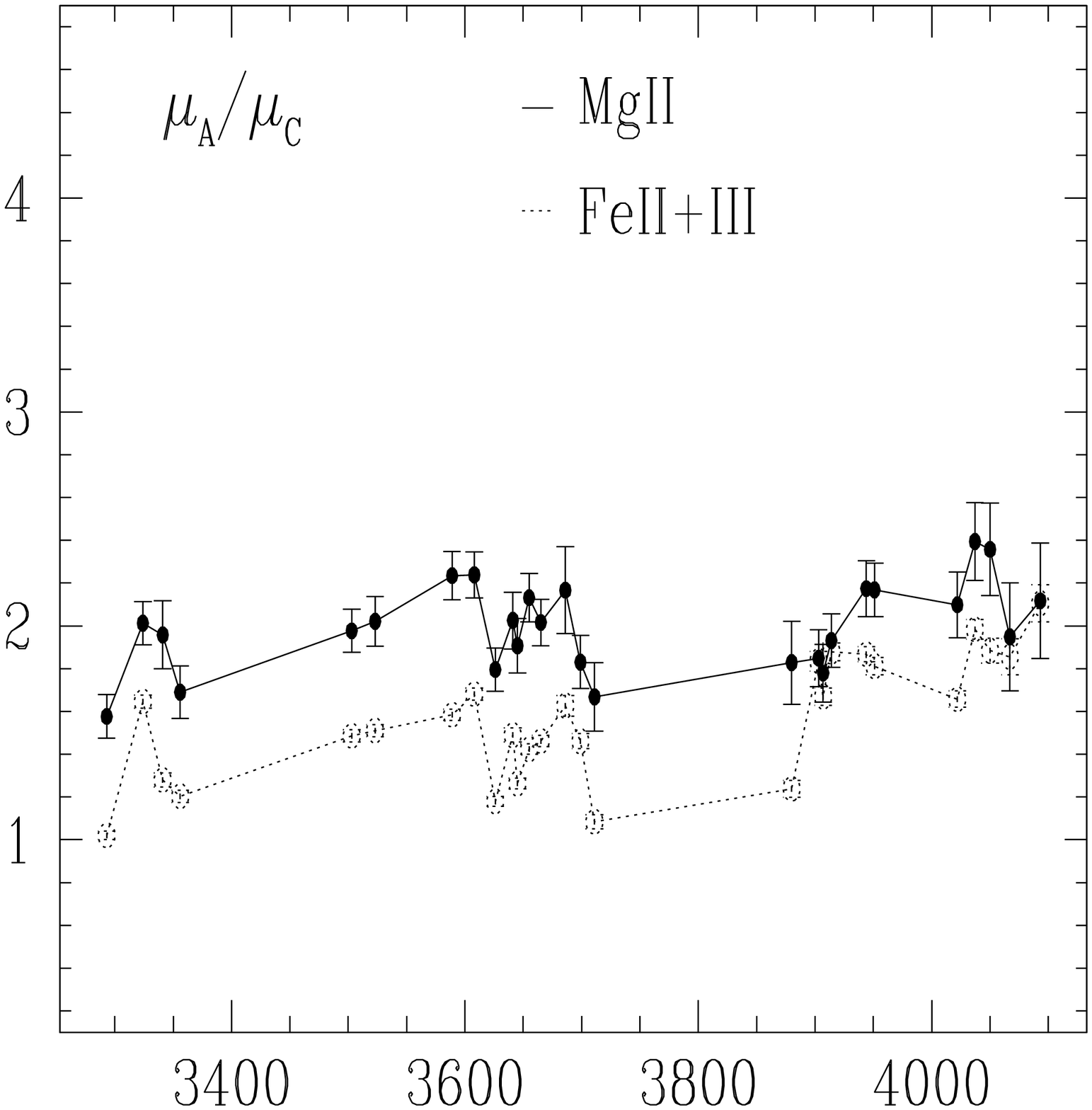}\\
\includegraphics[height=3.6cm,width=5.1cm]{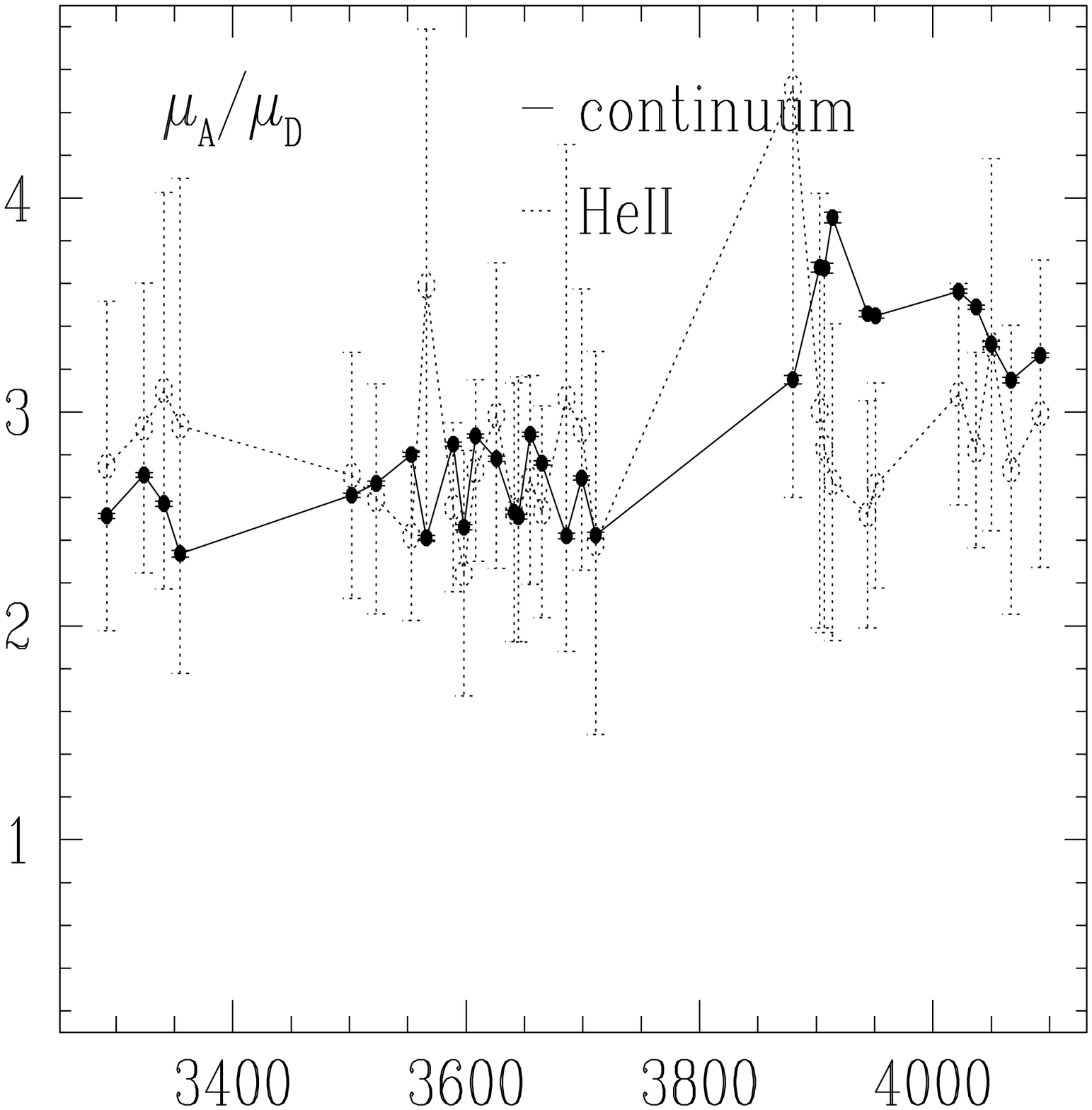}& 
\includegraphics[height=3.6cm,width=5.1cm]{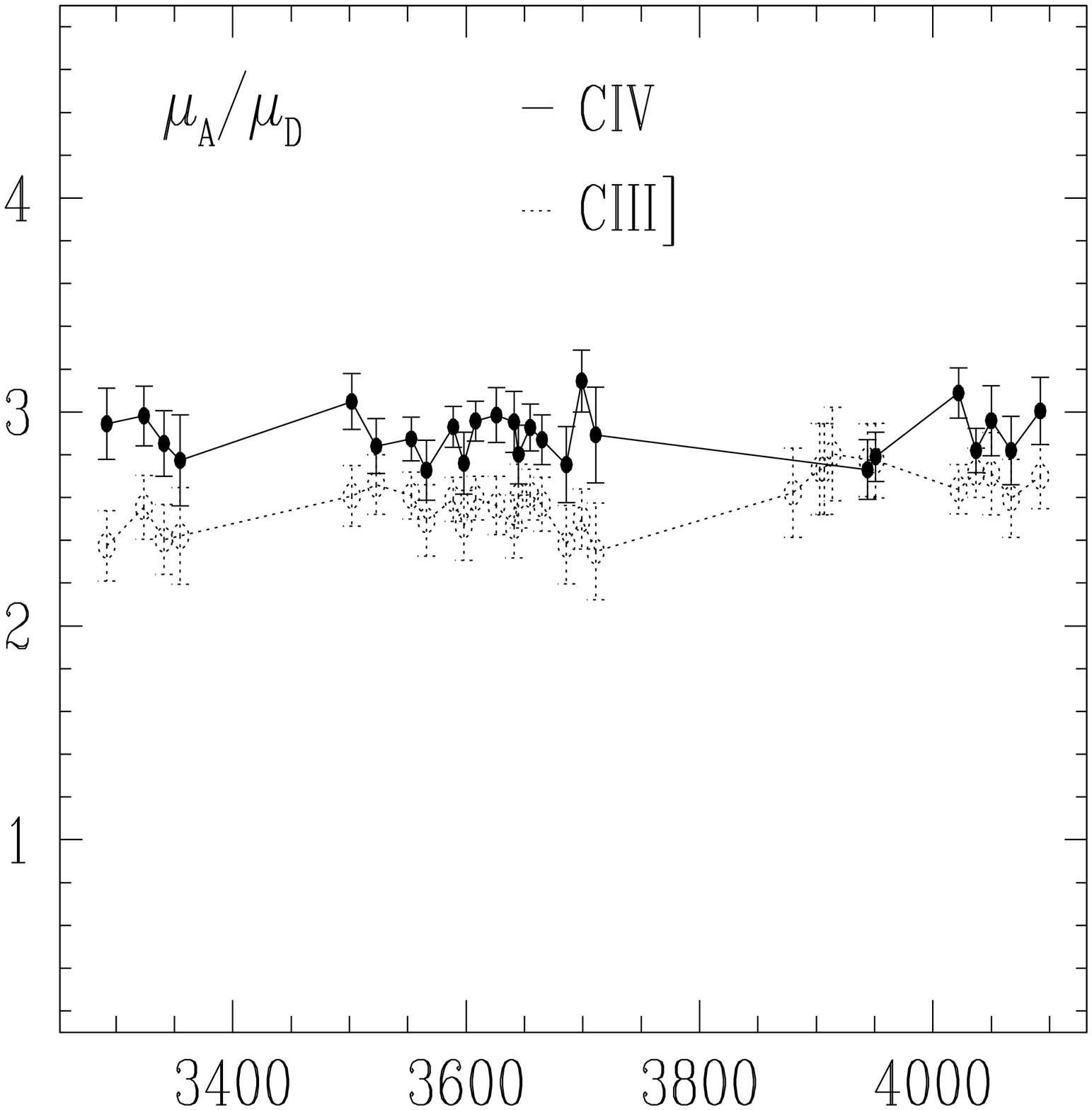}& 
\includegraphics[height=3.6cm,width=5.1cm]{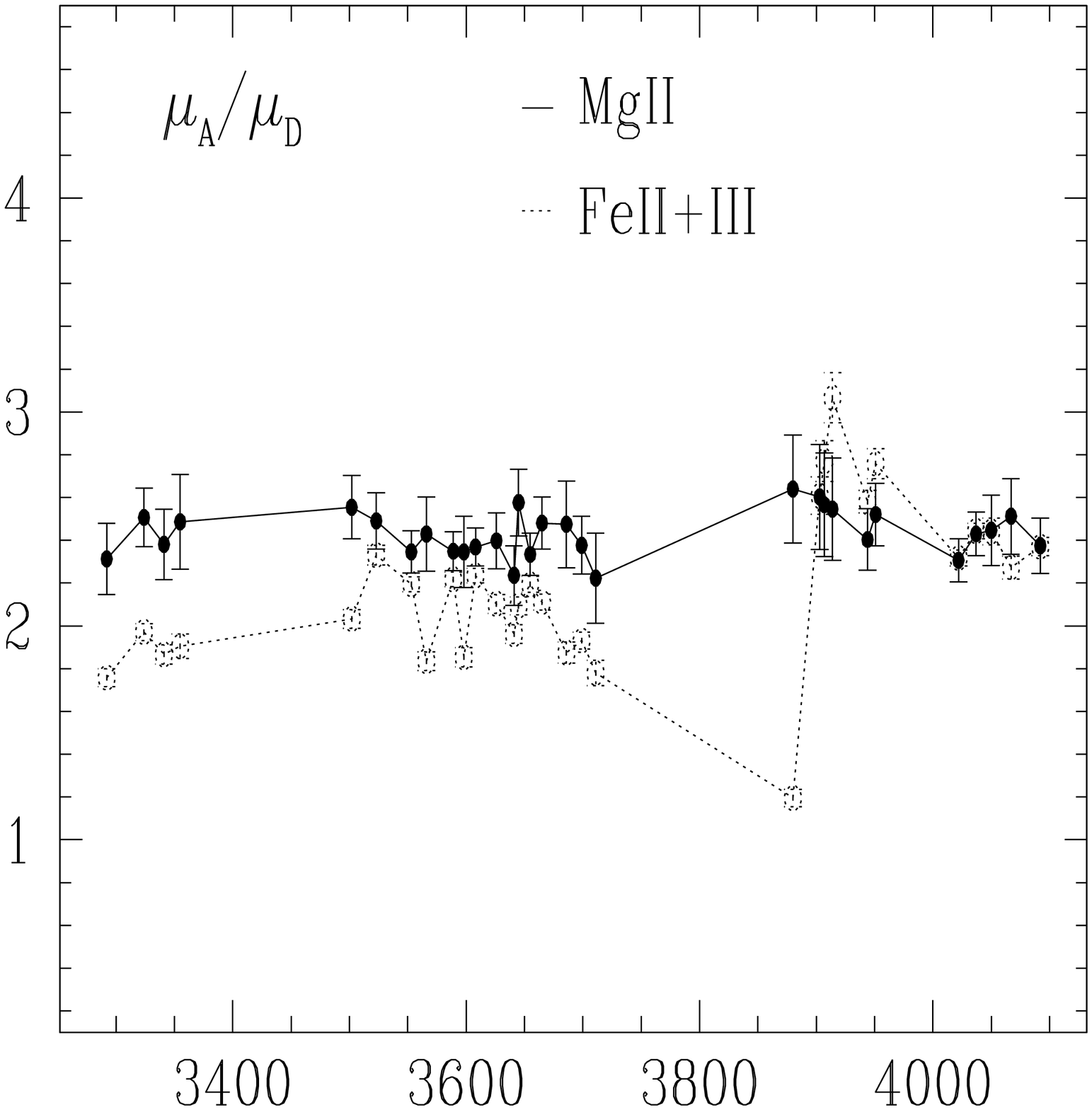}\\
\includegraphics[height=3.6cm,width=5.1cm]{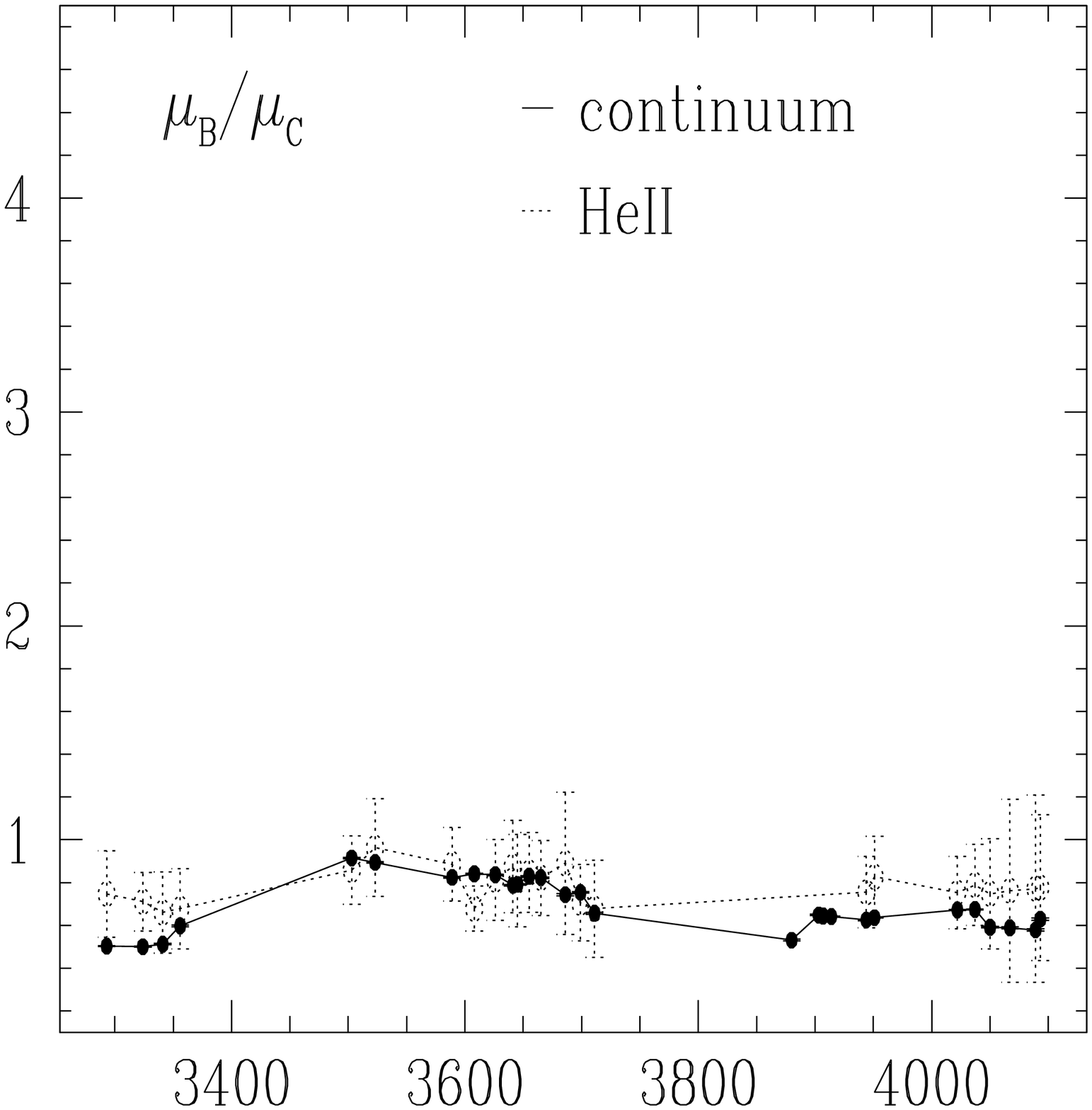}& 
\includegraphics[height=3.6cm,width=5.1cm]{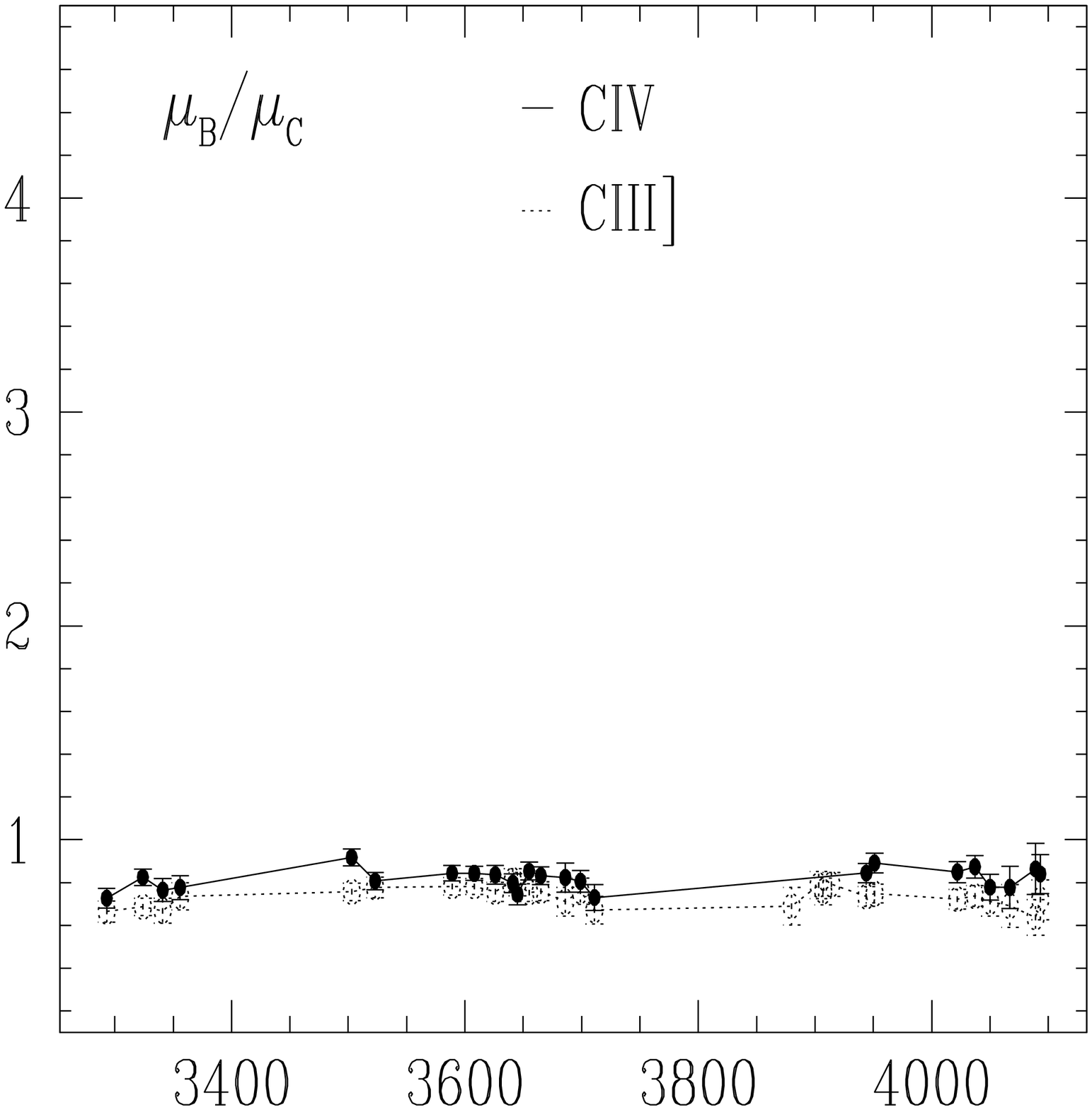}& 
\includegraphics[height=3.6cm,width=5.1cm]{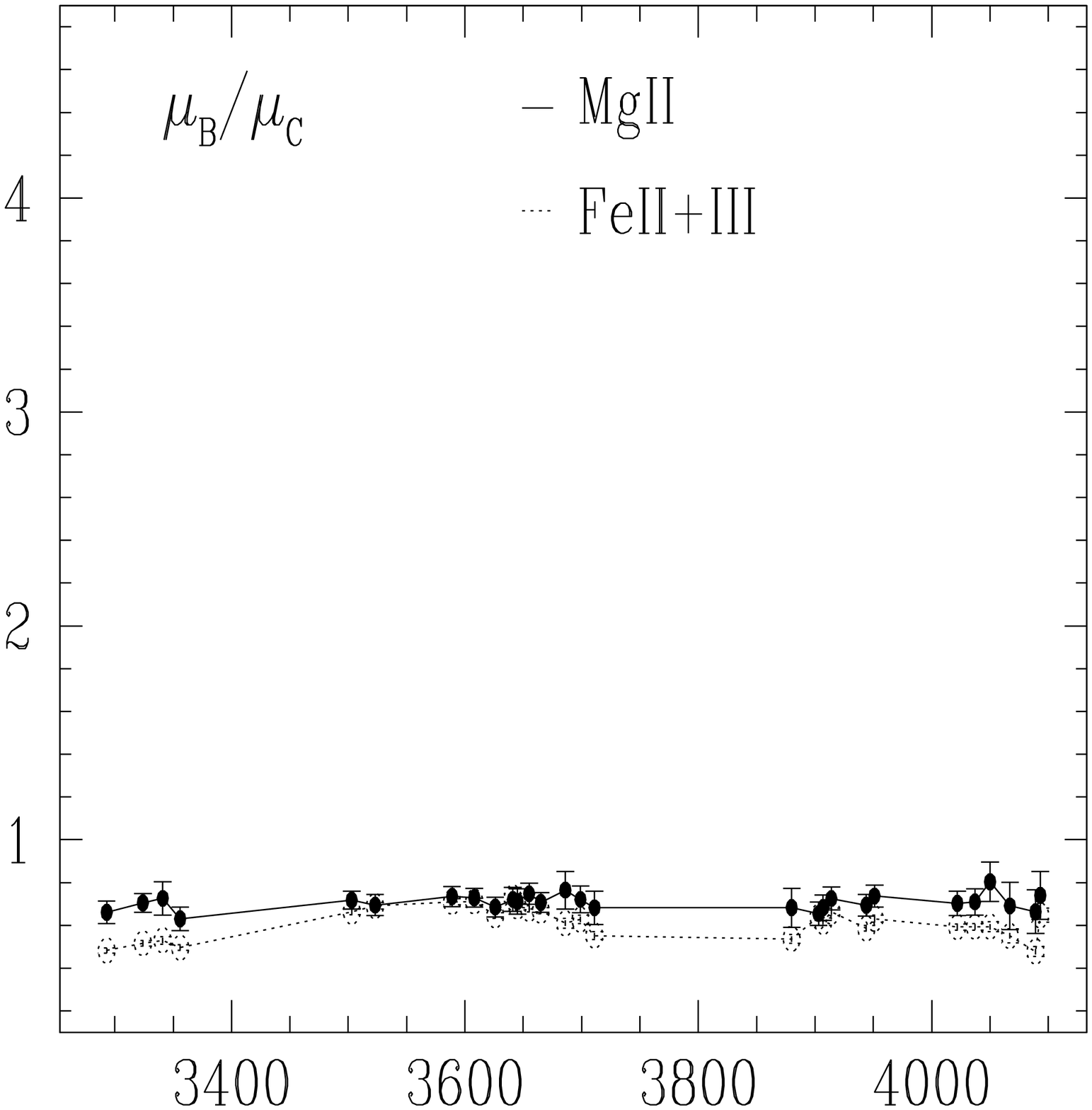}\\
\includegraphics[height=3.6cm,width=5.1cm]{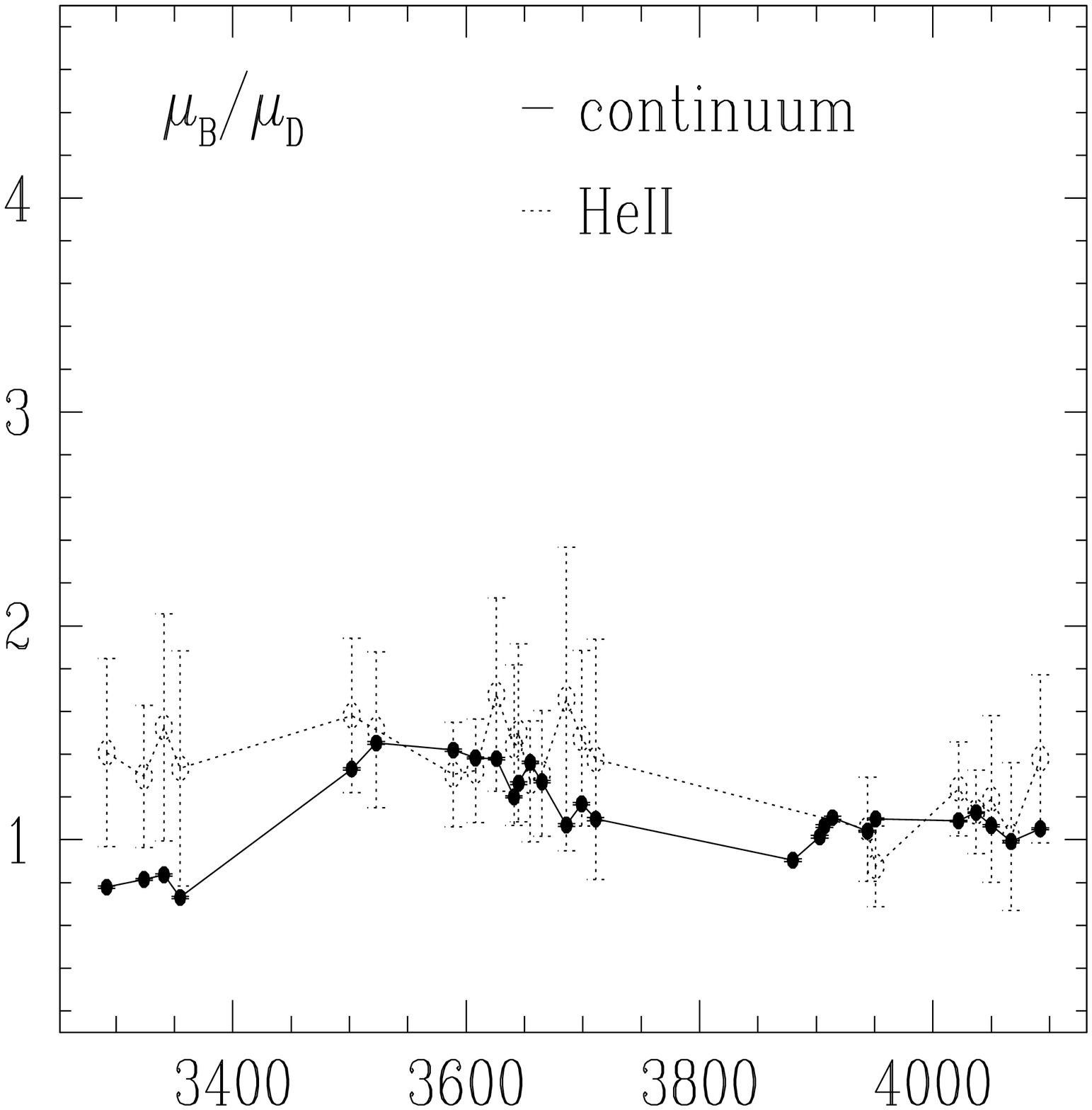}& 
\includegraphics[height=3.6cm,width=5.1cm]{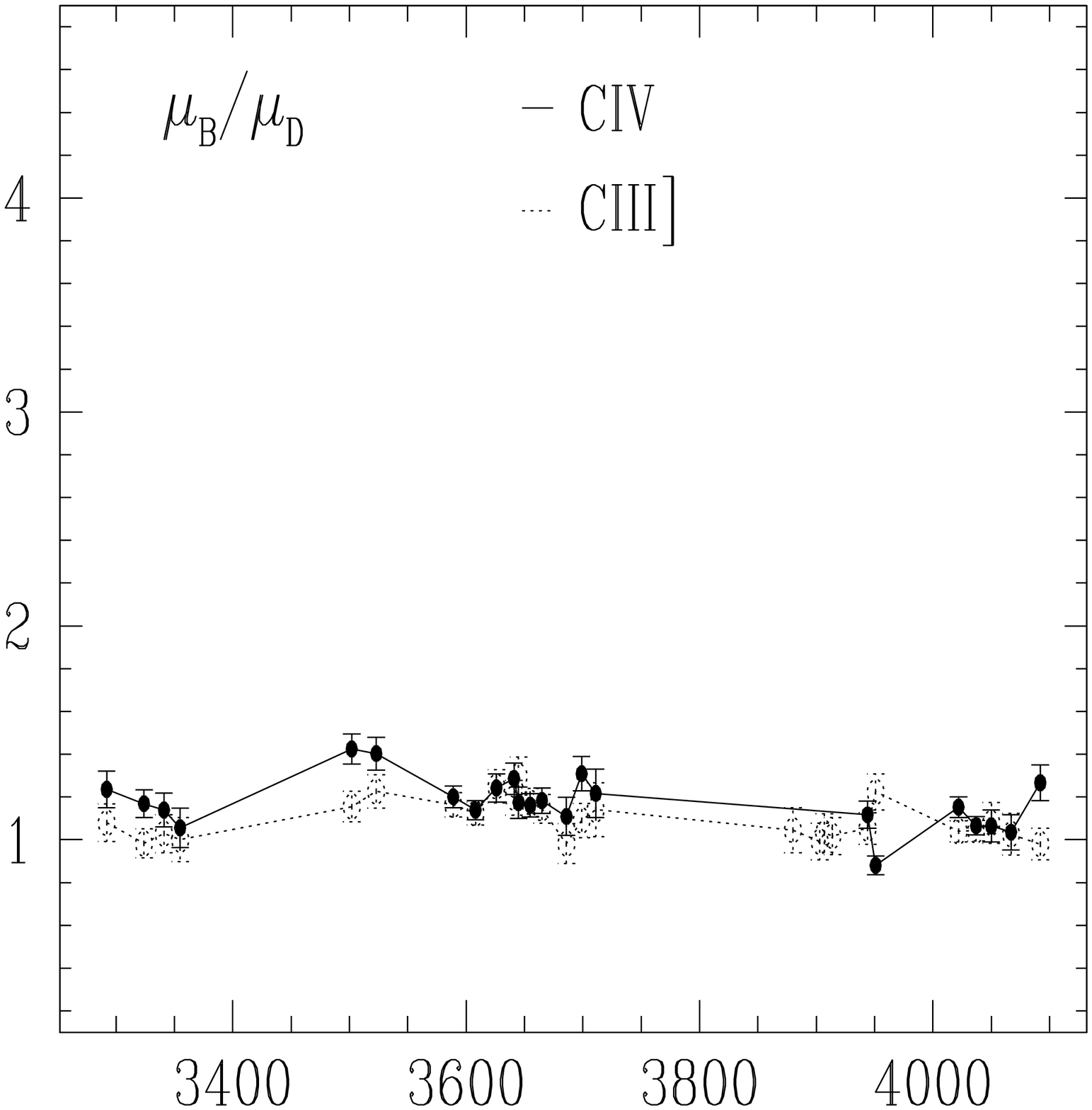}& 
\includegraphics[height=3.6cm,width=5.1cm]{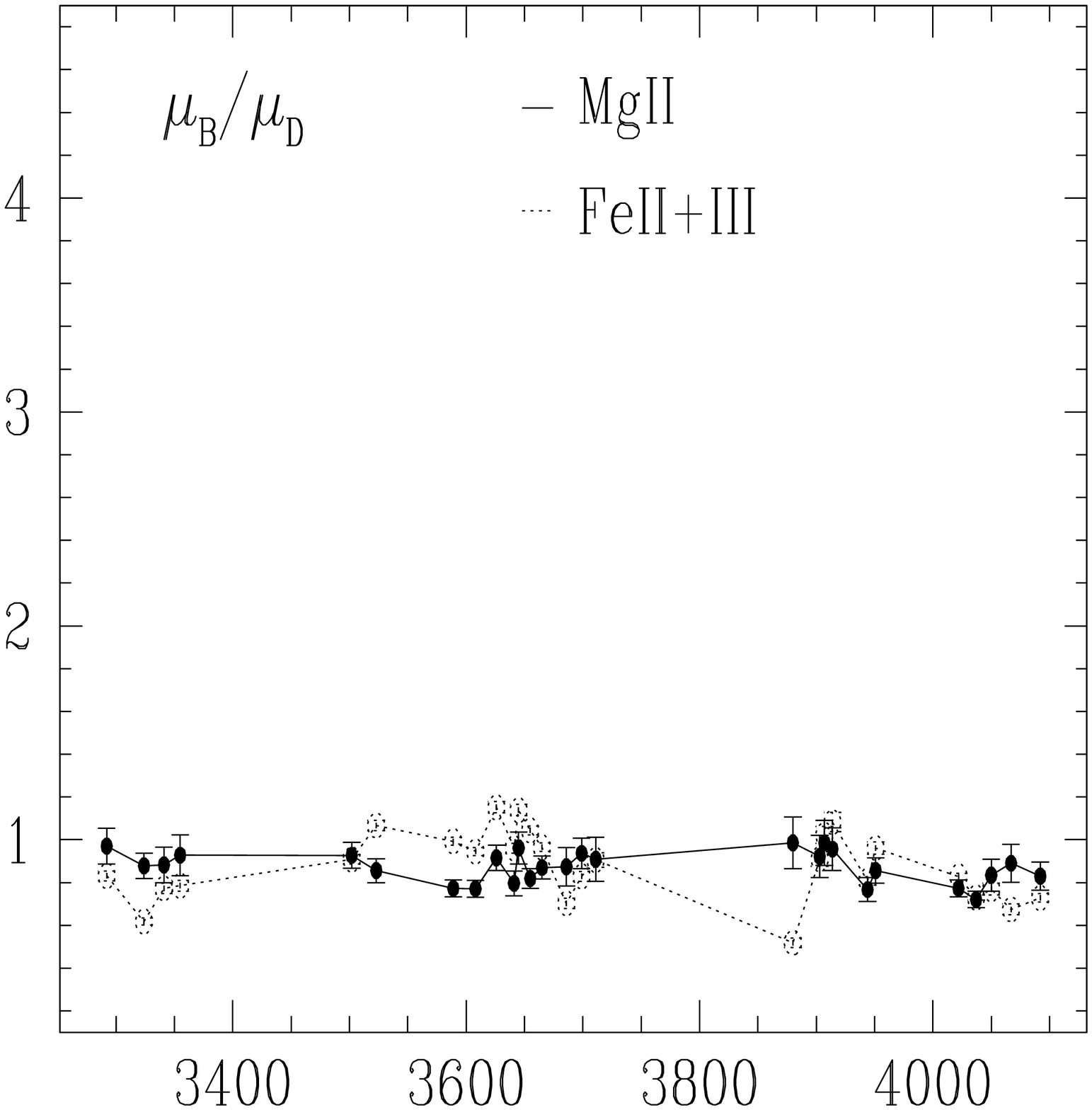}\\
\includegraphics[height=3.6cm,width=5.1cm]{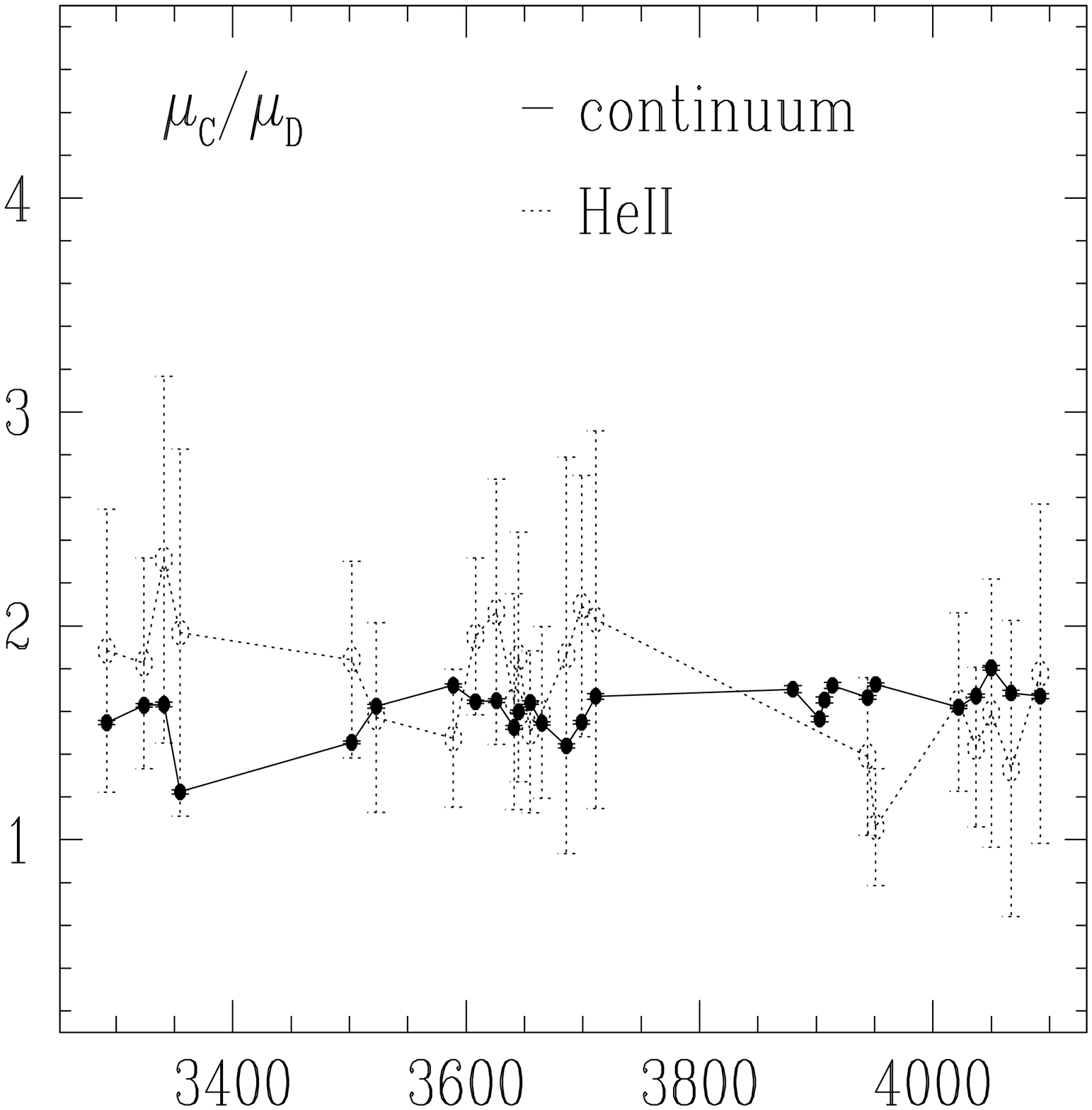}& 
\includegraphics[height=3.6cm,width=5.1cm]{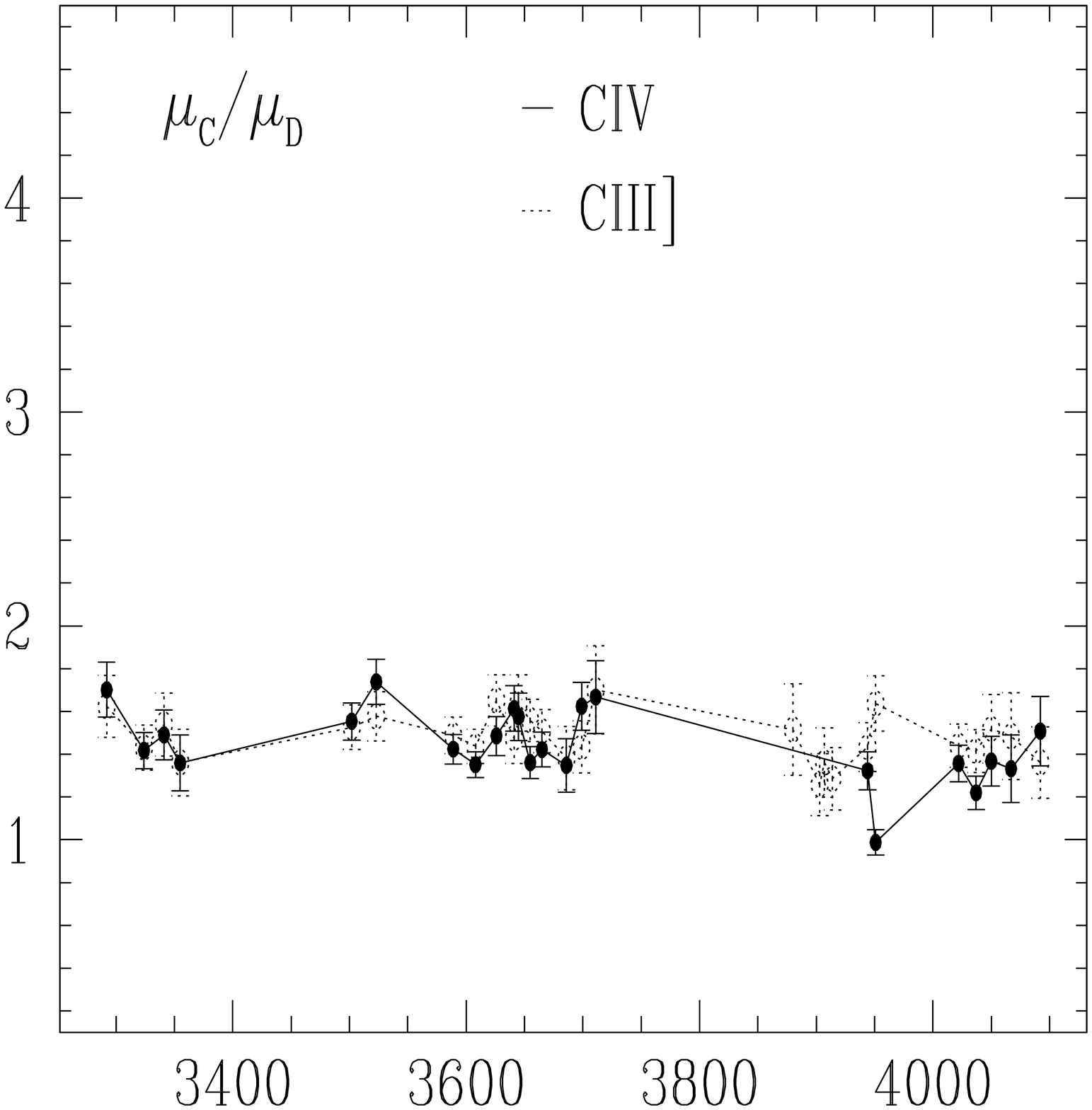}& 
\includegraphics[height=3.6cm,width=5.1cm]{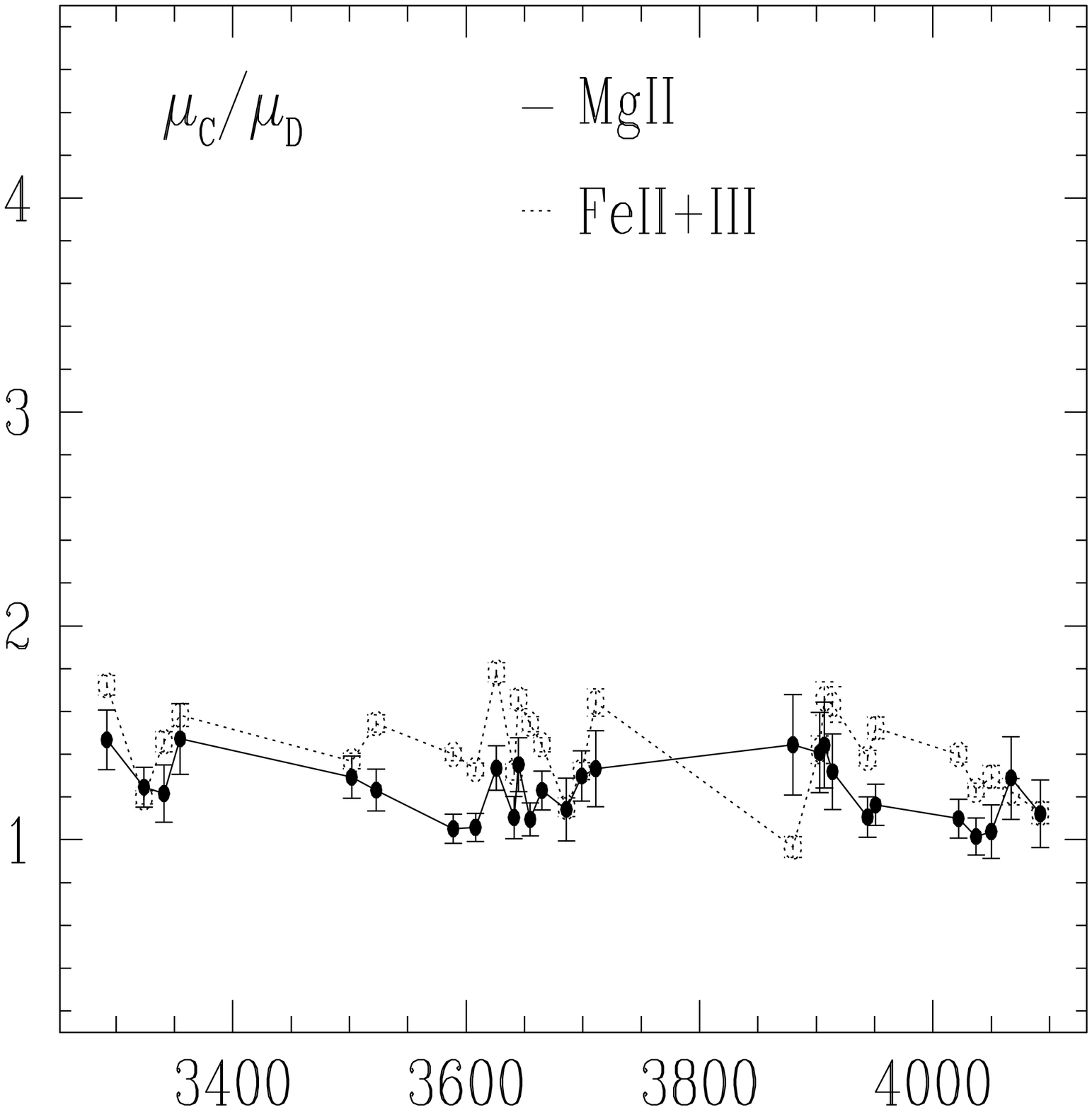}\\
\end{tabular}
\caption{Microlensing-magnification
ratios  $r_{ij}$  as a function of time  (HJD$-2450000$)
for all possible  combinations   of image pairs. 
These ratios are corrected for the  macromagnification by the lensing galaxy
following equation  (\ref{mumu_eq}).  The ratios are given
 for the   continuum, for  the
BELs \ion{He}{II}, \ion{C}{IV}, \ion{C}{III]}, \ion{Mg}{II} and for the iron
pseudo-continuum.}
\label{mumu}
\end{center}
\end{figure*}

In Fig.~\ref{mumu}, we  show   the variations of $r_{ij}(t)$  for   the
integrated flux in  the main emission lines and  in  the continuum. We
plot in the figure all  possible combinations of image ratios. Several
interesting results can be drawn from this figure:

\begin{enumerate} 

\item With the exception of $r_{BD}$ during the last season
(i.e. HJD~$>$~3900 days), none of the ratios  $r_{ij}$ are close to
1,  meaning that microlensing is acting  at least  in three quasar
images during the entire  monitoring. The evidence for $r_{BD}\sim$
1 when HJD~$>$~3900 days supports the absence of significant microlensing
in images B and D during the last observing season.

\item All the BELs have $r_{ij}\neq 1$ (except $r_{BD}$).  
The $r_{ij}$ ratios  in the BELs  generally closely follow  the  value
measured  for   the continuum.   This  demonstrates that the BLR is small
enough - probably not larger than a few Einstein radii of a typical microlens
in \obj\ - so that the BELs can be significantly affected.
In addition, the  variations observed
in the BELs are correlated with those in the continuum.

\item The largest changes of magnification
ratios   involve  images  A  (for HJD~$\sim$~3900    days)  and B (for
HJD~$\sim$~3500 days). This is  seen already in  Fig.~\ref{micro}, but
the ratios shown here allow us to make sure that the variations are
only due to microlensing.

\item The ratios $r_{Aj}$ and $r_{Cj}$ deviate significantly from 1 during the
whole  observing period. As an example,  we measure in the continuum
(resp.  \ion{C}{III]})  $r_{AD} \sim 2.6$  (resp. 2.8)  for HJD~$<$~3900
days and $r_{CD} \sim 1.6$  (resp. 1.6). Interestingly, these  ratios
have remained roughly the same since the beginning  of the OGLE-III campaign
(i.e.  HJD~$\sim$~2000 days).  Our values of $r_{AD}$ and $r_{CD}$ are
also similar to the value  measured in the \ion{H}{$\beta$} emission
line   in  August 2002    (HJD~$=$~2496	days)  by Metcalf  et	 al.
(\cite{met04}).  This strongly  suggests  that  images  A and  C are
affected by long-term micro/milli-lensing  on periods longer  than 5
years.

\item The ratio $r_{CD}$ is the most stable ratio along the monitoring
campaign, indicating that no major (short) microlensing event
occured in images C or D.

\end{enumerate}

Because of (b)  and (e), we can safely  consider that 
during the time span of our observations, image D is  the
less affected by  microlensing, both  on  short 
(i.e. of the order of a few weeks) and  
long timescales (years).  This is consistent with the
broad-band microlensing light curves of Fig.~\ref{micro}.

We use  image D as  a reference  to study  the ``short''  microlensing
events  affecting  image   A at  HJD~$\sim$~3900 days  and  image  B at
HJD~$\sim$~3500 days.  During  both  events the emission  lines  are not
magnified by the same  amount (Fig.~\ref{mumu}).  

To quantify
this,  we   compute in Table~\ref{mu_ratios}  the  mean  values of 
$r_{AD}$  and $r_{BD}$ during microlensing   events and in more
``quiescent'' phases.
For image A (i.e.  $r_{AD}$), we  clearly see in this table
that  \ion{C}{IV}, \ion{C}{III]}  and  \ion{He}{II}  show very similar
magnification ratios, while the
\ion{Mg}{II} line is less magnified.   

All the lines are less magnified than the continuum, consistent with a
scheme where  the continuum is emitted  in the most compact region, and
where other emission lines are  emitted in larger regions, the largest
region  being  the one   with  the lowest ionization  potential (i.e.
\ion{Mg}{II}). 
Indeed the ionization potentials of the different
lines are 47.9 eV (\ion{C}{IV}), 24.6 eV (\ion{He}{II}),
24.4 eV (\ion{C}{III]}), and 7.6 eV (\ion{Mg}{II}).
The difference of magnification between
\ion{C}{III]}  and  \ion{Mg}{II}  was  not   observed  by Wayth et
al. (\cite{wayth}).  For image B,  the same  global trend is observed
except that the relative errors on the  $r_{BD}$ ratios are higher due
to the lower signal-to-noise ratio of the spectra of image B. The effect
is almost absent when $r_{AD}$  and $r_{BD}$ are computed in 
quiescent phases of components A and B.

The behaviour of the \ion{Fe}{II+III} emission is
more difficult to interpret, as this complex is in fact a blend of many
lines. However, we note that the \ion{Fe}{II+III} complex in image A
is microlensed  at about  the same level  as the \ion{Mg}{II} line.  In
addition,  the  difference  in  magnification of  the \ion{Fe}{II+III}
lines between a microlensing and a quiet  phase is larger than for the
other   lines.  This  may suggest  differential   magnification of the
emitting regions within the \ion{Fe}{II+III} complex, i.e. that the
\ion{Fe}{II+III} is present both in compact and more extended regions,
a conclusion also reached by Sluse et al. (\cite{sluse}).

\begin{table}[t!]
\caption[]{Mean microlensing ratios for the continuum and for the
main BELs.   The mean  values for $r_{AD}=\mu_A/\mu_D$   and
$r_{BD}=\mu_B/\mu_D$ are computed for the observations around the epoch
in  the HJD  line,  i.e. during   microlensing events or  during
quiescent  phases. The values are given  along with  the dispersion of
the points around the mean. 
}
\label{mu_ratios}
\begin{flushleft}
\begin{tabular}{lcccc}
\hline
\hline
              	 &  $<r_{AD}>$      &  $<r_{AD}>$	&  $<r_{BD}>$	   &  $<r_{BD}>$ \\
\hline  HJD   	 &  3900 d	    &  3500  d  	&     3500  d	   & 3300 d \\
\hline  State 	 &   Micro-A	    &  Quiet-A  	&  Micro-B	   & Quiet-B \\
\hline
Cont.         	 &  3.46 $\pm$ 0.24 &  2.65 $\pm$ 0.18  &  1.28 $\pm$ 0.12 &  0.79 $\pm$ 0.05\\
\ion{C}{IV}   	 &  2.89 $\pm$ 0.13 &  2.90 $\pm$ 0.11  &  1.22 $\pm$ 0.11 &  1.15 $\pm$ 0.07\\
\ion{He}{II}  	 &  3.01 $\pm$ 0.55 &  2.70 $\pm$ 0.33  &  1.50 $\pm$ 0.23 &  1.39 $\pm$ 0.10\\
\ion{C}{III]} 	 &  2.71 $\pm$ 0.07 &  2.54 $\pm$ 0.09  &  1.17 $\pm$ 0.09 &  1.02 $\pm$ 0.04\\
\ion{Mg}{II}  	 &  2.49 $\pm$ 0.10 &  2.40 $\pm$ 0.10  &  0.88 $\pm$ 0.09 &  0.91 $\pm$ 0.04\\
\ion{Fe}{II+III} &  2.56 $\pm$ 0.23 &  2.05 $\pm$ 0.17  &  0.99 $\pm$ 0.15 &  0.75 $\pm$ 0.09\\
\hline  \end{tabular}  \end{flushleft}
\end{table} 

\begin{figure*}[t!]
\begin{center}
\includegraphics[width=8cm]{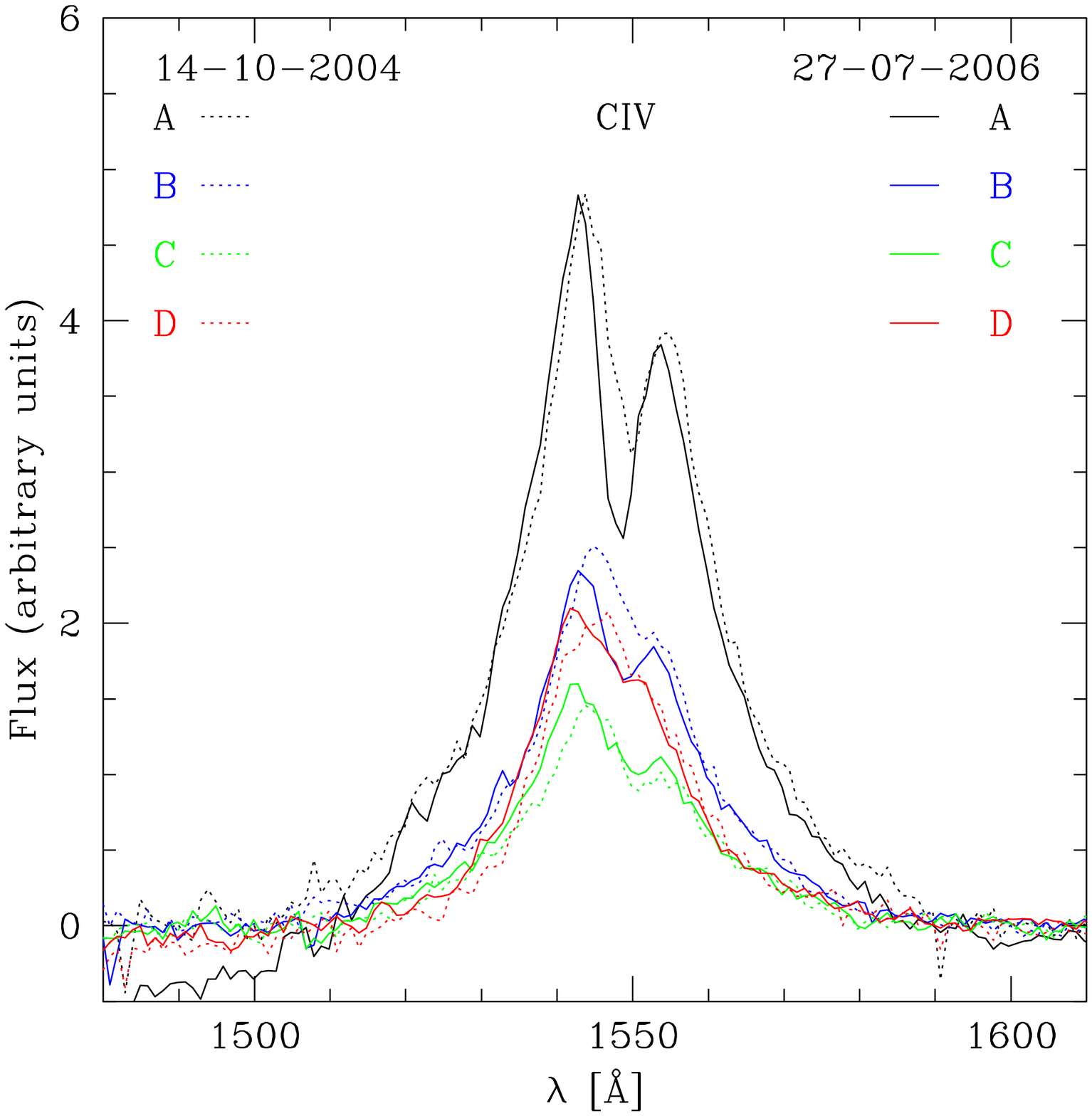}
\includegraphics[width=8cm]{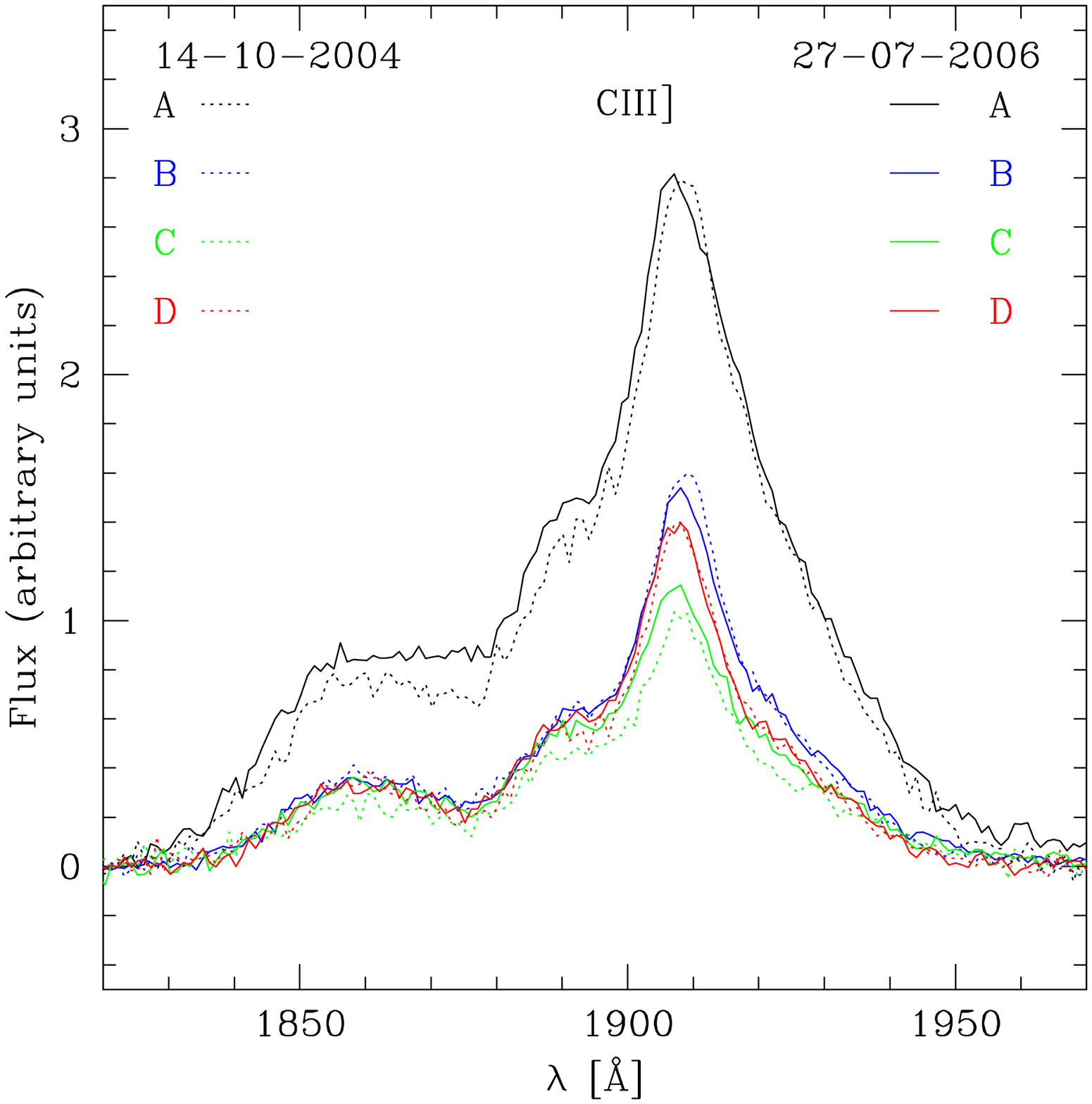}
\end{center}
\caption{The  \ion{C}{IV} (left) and \ion{C}{III]} (right)
    BEL profiles for quasar images A, B, C and D as observed at 
    epoch \#~1 (14$-$10$-$2004, dotted) and epoch \#~24 (27$-$07$-$2006, solid). 
    The first epoch is used as a reference while the second
    epoch falls within the high-magnification
    episode that occured at HJD~$\sim$~3900 days in
    image A (Sect.~\ref{emline}). In order to properly show the
    extra-magnification of the line wings during the strong microlensing 
    episode of image A, we have multiplied
    the line profile in A observed on 14$-$10$-$2004 by a factor 1.05. }
\label{linewidth}
\end{figure*}

\begin{figure*}[p!]
\begin{center}
\includegraphics[width=7.5cm]{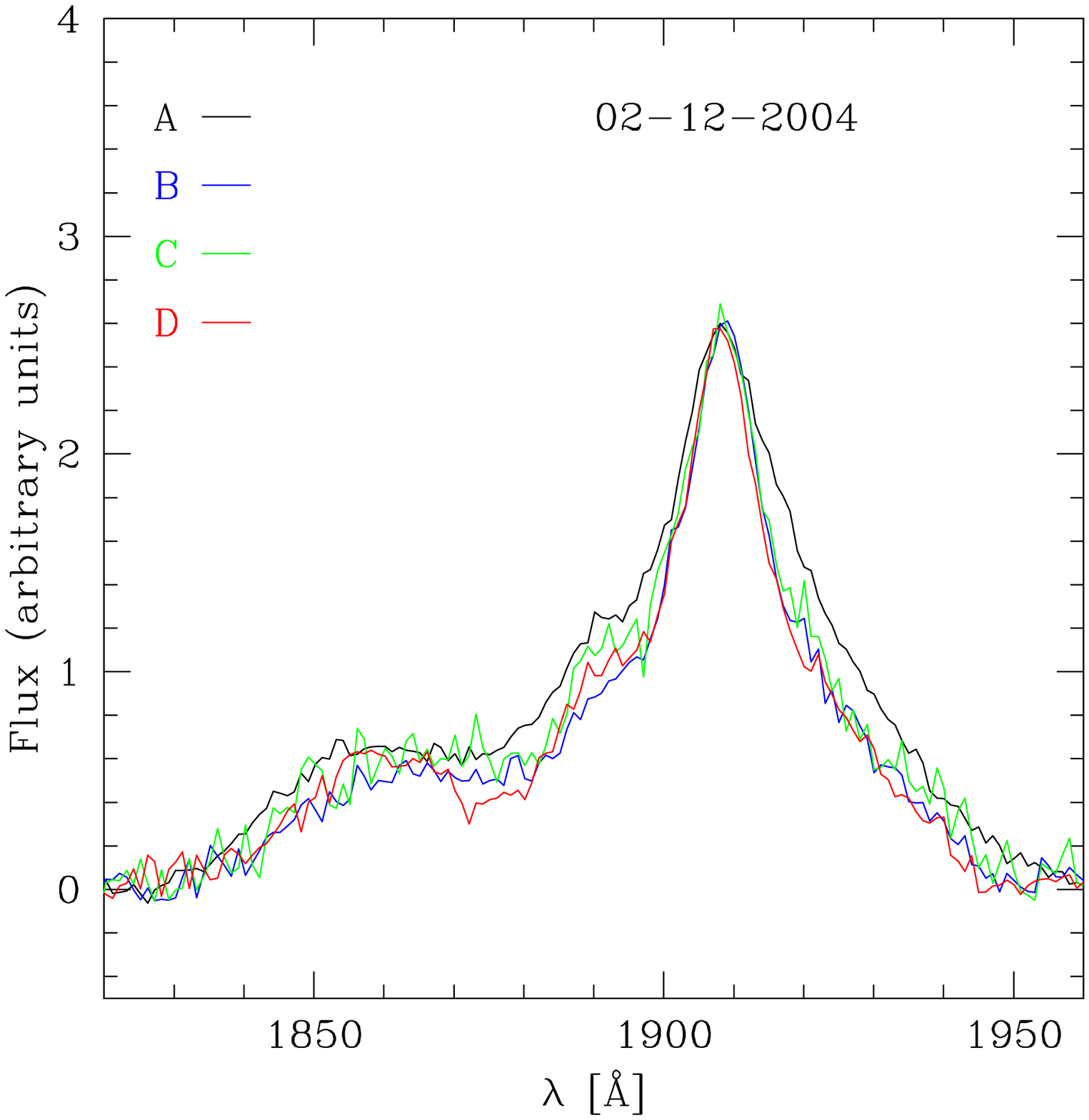}
\includegraphics[width=7.5cm]{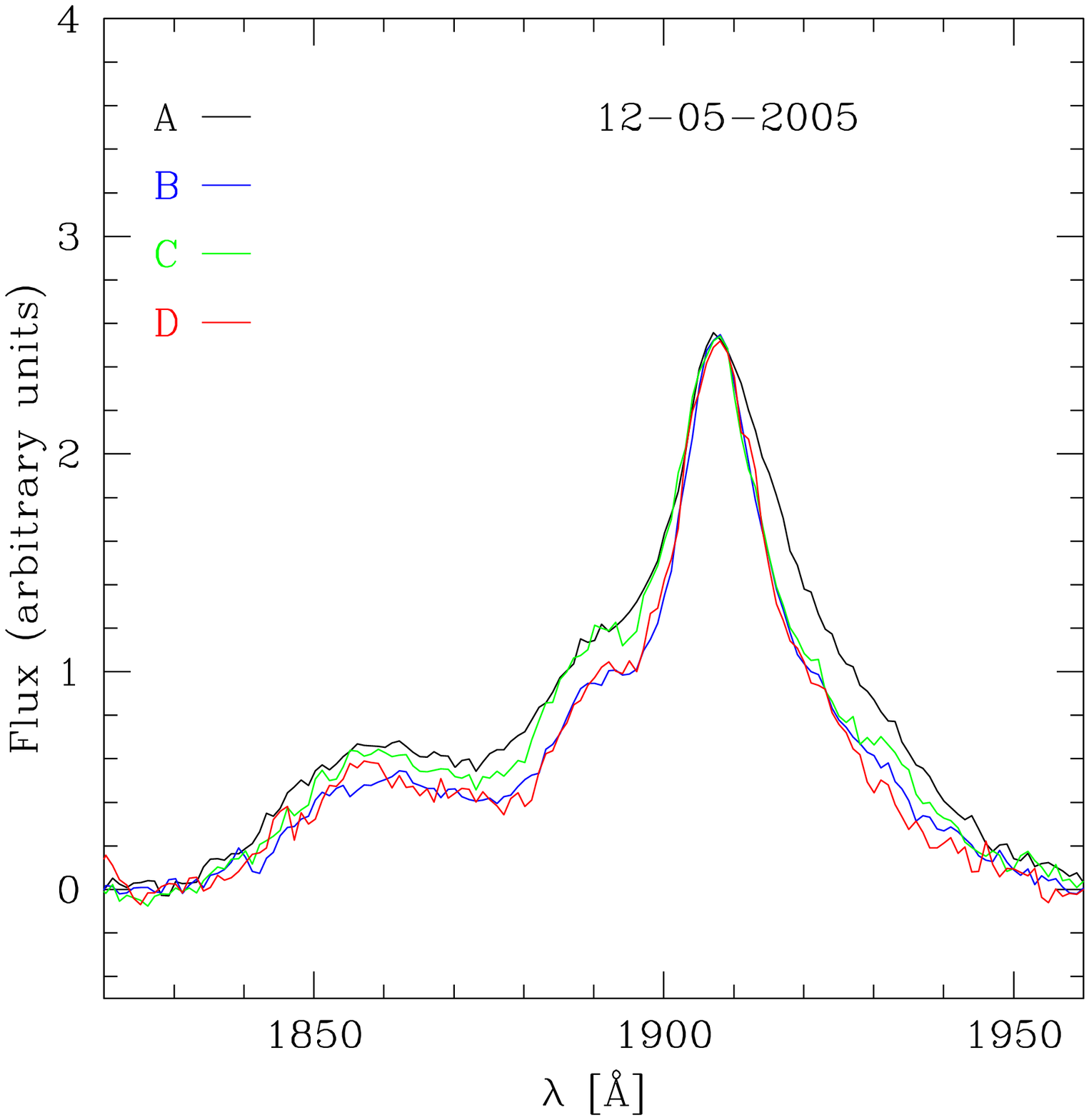}
\includegraphics[width=7.5cm]{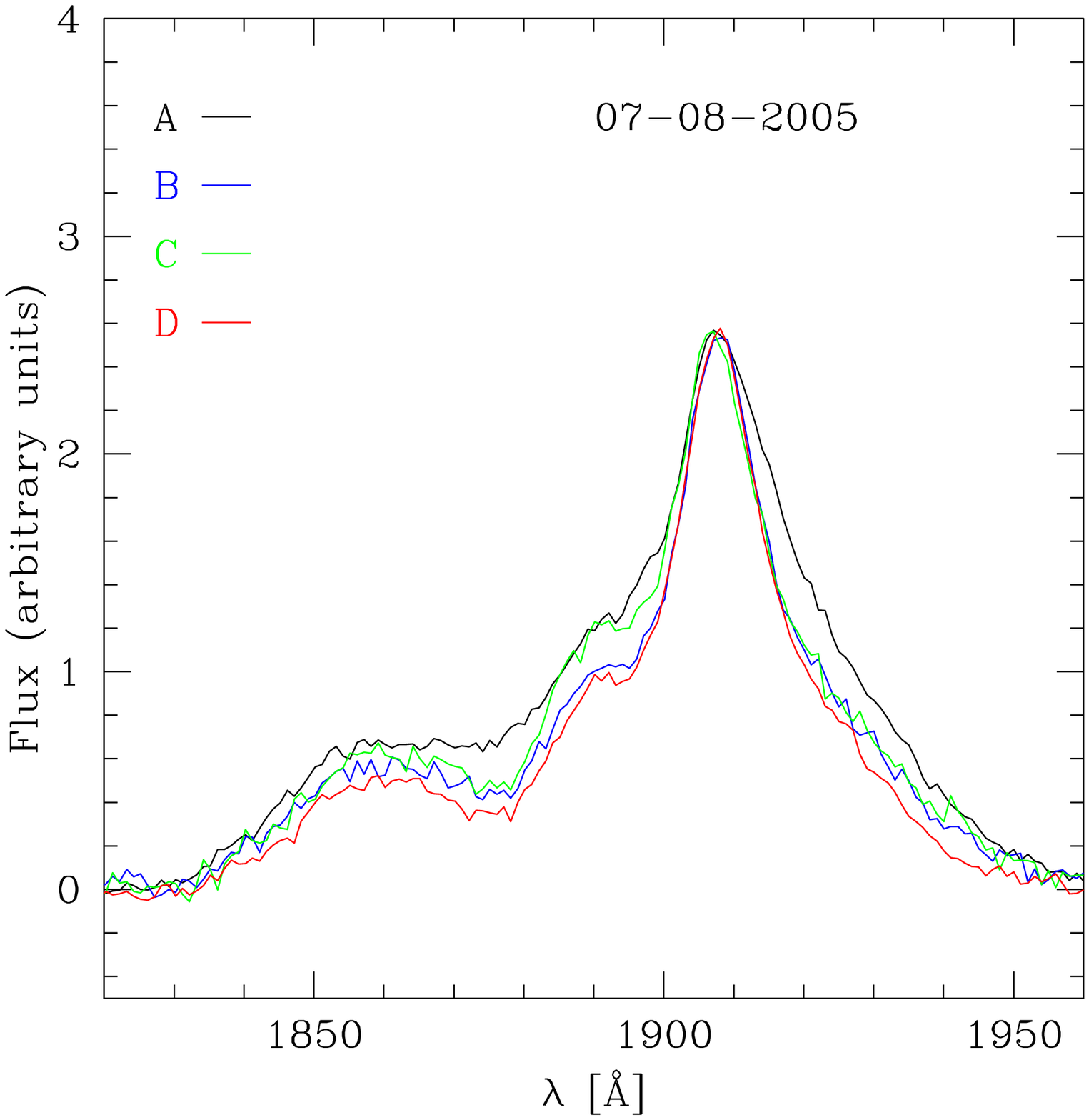}
\includegraphics[width=7.5cm]{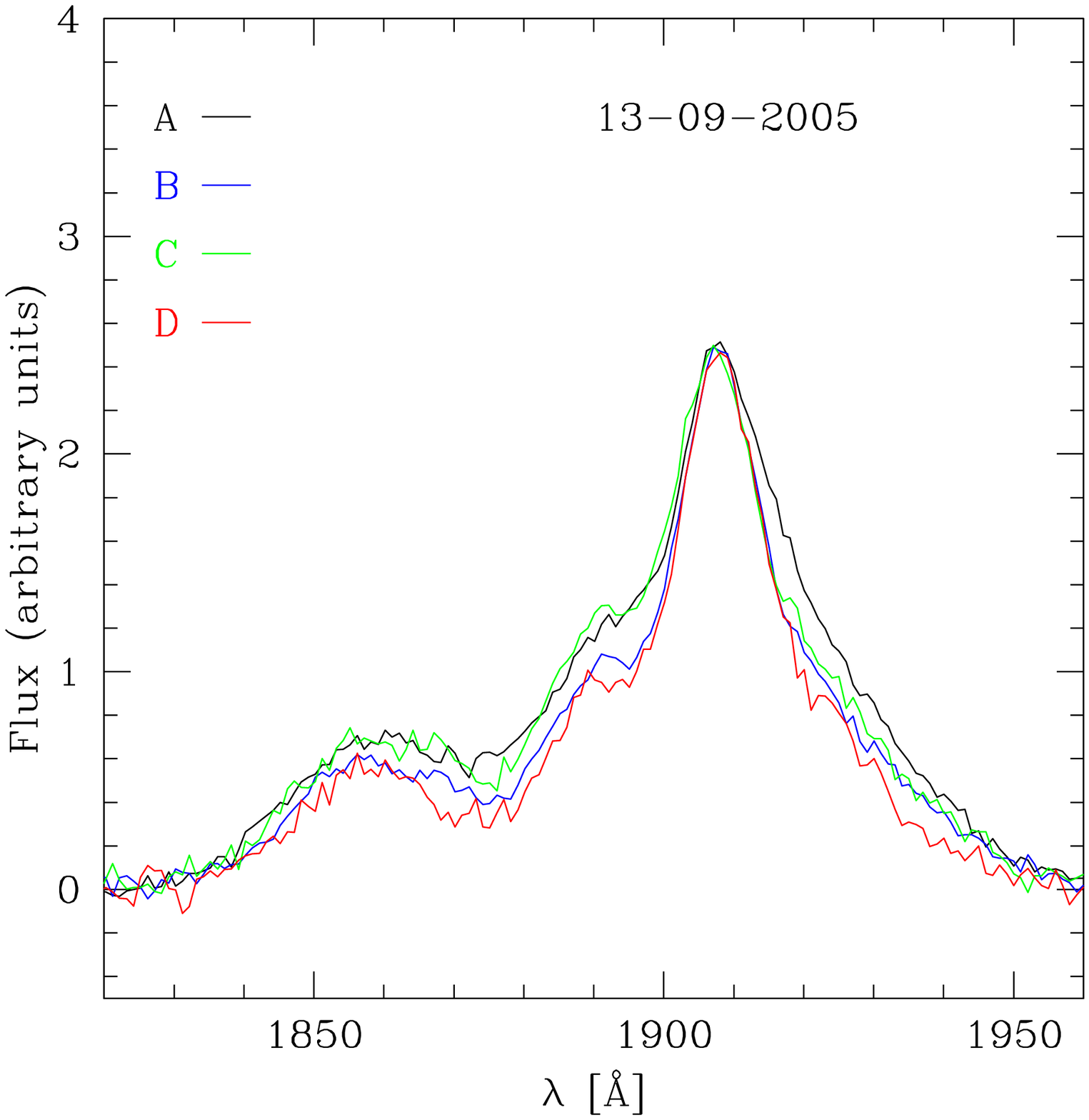}
\includegraphics[width=7.5cm]{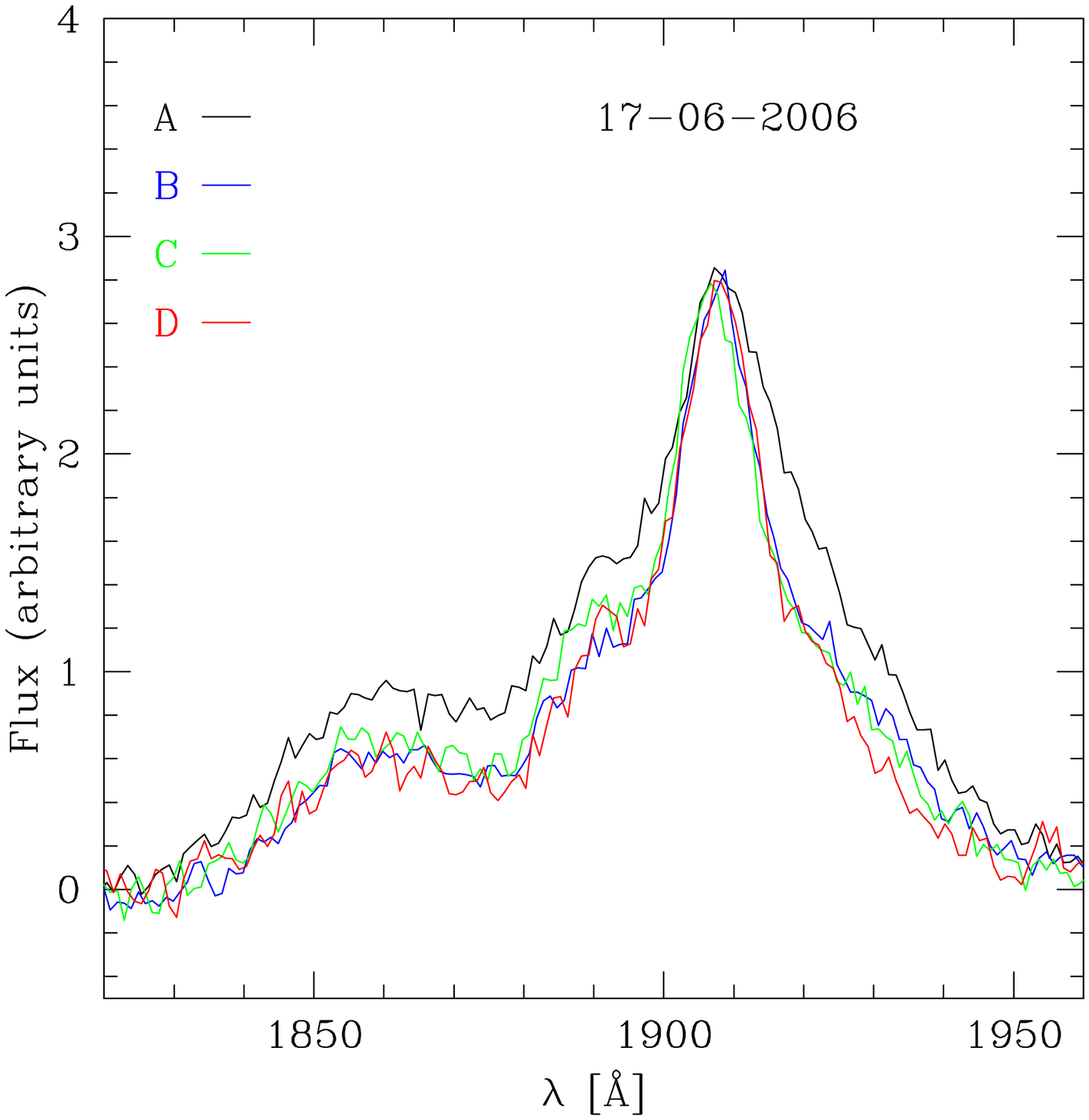}
\includegraphics[width=7.5cm]{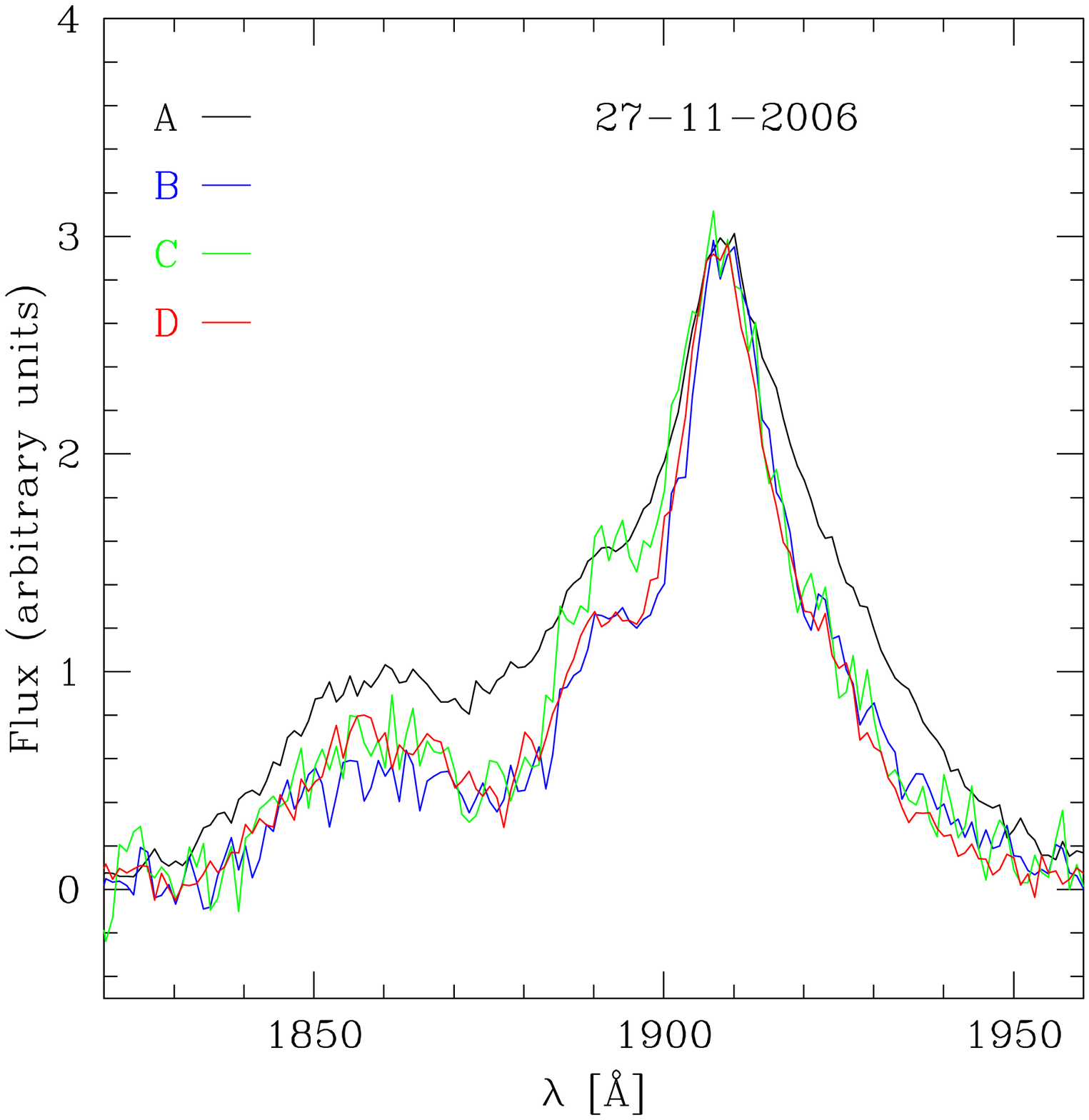}
\caption{Comparison of the \ion{C}{III]} BEL profiles of images A, B, C and D
    of the lensed quasar \obj\ on 6 different epochs. The line
    profiles are normalized so that they have the same peak
    value. This representation illustrates the difference in 
    the line profiles of the 4 lensed images at
    several epochs.  The most striking effect is the larger wings observed
    at all epochs in image A. These larger wings are caused by differential microlensing
    between the wings and the core of the \ion{C}{III]} line. Small microlensing
    induced fluctuations of the lines profiles are also observed in
    images B, C, and D. The epochs are chosen to sample the whole observing
    period.}
\label{Cline}
\end{center}
\end{figure*}

\subsection{Line profiles}
\label{emline}

We  have discussed in   the  previous  sections the  global  intensity
changes in  the  emission lines   of \obj.   We  now  investigate  the
possibility of a change in their profile.  Such profile variations may
be  caused by  differential  magnification of  regions with  different
velocities in the BLR.  This can introduce,  e.g., asymmetric changes or
even peak  displacement of the    line (Lewis \cite{lewis98a},  Abajas
et~al.\cite{abajas}, Lewis \& Ibata \cite{lewis}). Observational evidence
for profile variations has been reported, e.g. in 
HE~2149-27 (Burud et al. \cite{burud});
SDSS~1004$+$4112 (Richards et al.  \cite{richards}; G\'omez-\'Alvarez et al. \cite{gomez});
SDSS~J0924$+$0219 (Eigenbrod et al.  \cite{cosmograil2}); and  
RXJ~1131$-$1231 (Sluse et al.  \cite{sluse}). 

We study  the  variability of  the emission  lines in  the four quasar
images  using  the  continuum-subtracted spectra   obtained  from  the
decomposition  procedure of  Sec.~\ref{multicomp}.  
We concentrate  on the \ion{C}{III]}
emission line, which is easily decomposed  into a sum of two components
with full-width-half-maximum values about $1\,100$~\kms\ and 
$5\,000$~\kms. We will refer to these components in the following as
the ``narrow'' and ``broad'' components, although the difference  made
here between narrow and broad is only phenomenological.  In particular,
the  narrow  component is not associated, {\it  a  priori}, with the
narrow line region (NLR). 

We measure the microlensing magnification ratio
$\mu_{i}({\textrm{narrow}})/\mu_{i}({\textrm{broad}})$ for the
\ion{C}{III]} emission line in each quasar image.  In image A, the
magnification of the broad component is $\sim$1.8 times larger than
the magnification of the narrow component, and this does not change
drastically along our whole monitoring campaign.  In the
other quasar images, the microlensing-magnification of the narrow and
broad components are comparable. This implies that the
\ion{C}{III]} emission line is microlensed globally in all images,
except in A. In the latter image, the broadest part of the line is
more microlensed than the core, indicating that the broadest component
of the \ion{C}{III]} emission is emitted in a more compact region than
the core of the emission line.

The core of the narrow  \ion{C}{III]} emission line is probably 
associated mainly with photons emitted by the NLR: we observe
nearly the same  magnification ratio between  the narrow and the broad
components of the \ion{C}{III]} emission line  than is reported between the narrow
\ion{[O}{III]} and  the broad \ion{H}{$\beta$} emission lines (Metcalf
et al.~\cite{met04}). This gives a hint that 
the narrow component of the \ion{C}{III]} emission line may be partly
emitted by the NLR.

However, we do not measure the same microlensing-magnification ratios $r_{ij}$
in the narrow \ion{C}{III]} emission as those measured by 
Metcalf et al. (\cite{met04}) in the narrow \ion{[O}{III]} emission line.  
The $r_{BC}$, $r_{BD}$, $r_{CD}$ ratios measured by these authors 
are similar for the \ion{H}{$\beta$} BEL and for the narrow \ion{[O}{III]}
emission line, but are 
different from 1, which may be interpreted as a consequence of the microlensing of the NLR. 
If the NLR is indeed microlensed, the discrepancy between our 
values and the ones by Metcalf et
al. (\cite{met04}) can simply be explained  by  the fact that the amplitude of
microlensing changes with time. 
However, there are also two other possible explanations. 
First, 
the size of the
\ion{[O}{III]} emission region might be comparable with the size of the
macrocaustic of the lensing galaxy, such that the \ion{[O}{III]}
lensed images are extended and can be resolved by integral field spectroscopy
(Metcalf et al.  \cite{met04}, Yonehara
\cite{yon06}). This may lead to uncertainties
in the flux measurements depending on the chosen size of the aperture.
Finally, the discrepancy may be explained as well in
terms of extinction by the lens (see Sec.~\ref{extinction}).

The details of the \ion{C}{III]} profile also show variations. Our
spectra show evidence for systematic broadening, by $\sim$10\% of the
\ion{C}{III]} emission line in quasar image A  during our last
observing season (HJD~$>$~3900 days; Fig.~\ref{linewidth}).
In addition, we note evidence for variations in the central parts of
the \ion{C}{IV} emission   lines in  all  components and at
most  epochs.  An example  of   these variations is
shown in Fig.~\ref{linewidth}. Their interpretation will be much more 
complex than for the \ion{C}{III]} due to the decomposition of the line
into three emission components plus an absorber-like feature. 

To look for distortions in the line profiles of
images B, C, and D, we normalize the continuum-subtracted spectra of all
images so that they share the same \ion{C}{III]} central
intensity. The result is shown for selected epochs in
Fig.~\ref{Cline}.  With this choice of normalization and in the
absence of microlensing, 
the \ion{C}{III]} lines at a single epoch would match perfectly. 
It is conspicuous in Fig.\ref{Cline} that this is not
the case.  First, we clearly see the effect of differential
magnification between the core and the wings of image A at all epochs
(the wings of \ion{C}{III]} in image A are always larger than in the
other emission lines). The emission lines in B, C, and D are more similar to
one another but there is no epoch where the three line profiles match
perfectly, indicative that small microlensing fluctuations affect the
\ion{C}{III]} line. Image C also shows line profile variations, 
even though the effect is less pronounced than in image A. 
Finally, we note the absence of any strong line profile variations 
of image B during the short-term microlensing event occuring in
that image at HJD~$\sim$~3500 days.



\subsection{Differential extinction by the lensing galaxy}
\label{extinction}

Most of the magnification  ratios $r_{ij}$  fluctuate around a  mean
value.   The  fluctuations themselves    can   only be explained    by
microlensing, but the  value of the mean $<r_{ij}>$ is usually different from 1
during our observations.  This  can, in principle, be
explained either    by  long-term microlensing   or  by non variable
extinction by dust in the lens.

In order to test the latter hypothesis,  we use a simple and empirical
diagnostic using ratio spectra of  pairs of quasar  images.
In the absence  of reddening and  microlensing, these ratio spectra should be
flat.  If dust is present in different amounts on the lines of sight of
the images,  the ratio spectra will show  a non-zero slope, constant with time.
Any  time-variable   change  of slope can    safely  be  attributed to
microlensing.

We find that the most useful pairs are the ones  formed by A\&C, B\&C,
and C\&D.  Indeed, the C/D ratio  spectrum is found  to be almost flat
all  along    the  years,   indicating   no  significant  differential
extinction.  This is not  the case for the two  other pairs of ratio spectra
which  show reddening of image  C  relative to both  A  and B, as also
reported by Yee~(\cite{yee88}). We  estimate this differential reddening
using the extinction law by Cardelli et al.  (1989) and by assuming $R_V=3.1$.  We
find  that   the   differential  extinction  $A_V(C)-A_V(A)   \simeq
A_V(C)-A_V(B)$ is in the range 0.1-0.3~mag. This range  of values  is
sufficient to explain the discrepancy found between the $r_{ij}$ ratios  
measured in the \ion{C}{III]} narrow component and in  [\ion{O}{III}] of
Sec.~\ref{emline}.

Finally, our estimates of the extinction in the  lens are too small to
explain the highest values  of the mean magnification ratios observed
in Fig.~\ref{mumu} and Table~\ref{mu_ratios}. 
For instance, the mean of the $r_{AB}$ or $r_{AD}$ ratio 
reaches values larger than 2.
Static, long-term microlensing is therefore present in
the Einstein Cross, at least in images A and C.

\section{Conclusions}

This paper presents the first long-term (2.2 years) and  
well-sampled spectrophotometric monitoring of a gravitationally lensed quasar, 
namely the Einstein Cross \obj. 
The mean temporal sampling is of one observation every second week.
The observations are carried out with
the VLT in a novel way, using the spectra of PSF stars, both 
to deblend the quasar images from the lensing galaxy and to carry out
a very accurate flux calibration.  This paper, the first of a series, describes the
observations and the techniques used to extract the scientific
information from the data.

Detailed inverse ray-shooting  simulations will  be  needed  to infer
quantitative   information on  the internal  structure   of the lensed
quasar, and will be the topic of the future papers. The main observational
facts that  these simulations will need  to  take into  account can be
summarized as follows.

We find that all images of \obj\  are affected by microlensing both on
the long and short timescales.  Comparison  of the  image flux ratios  with
mid-IR measurements reveals  that  quasar images  A   and C  are  both
affected by long-term microlensing on a period longer than 5 years. This
long-term microlensing  affects  both   the  continuum  and the
BELs. 

Furthermore, in   quasar   image   A, the  broad   component   of  the
\ion{C}{III]} line is magnified by a factor 1.8 larger than the narrow
component.  On  the contrary, the   other quasar images have the  same
magnification in the narrow and broad components.

On  the short timescales, i.e. several months,  images A and  B are the most
affected by microlensing  during our monitoring campaign.  Image~C  and
especially  D are  the   most quiescent. Image~A  shows  an  important
brightening    episode  at    HJD~$\sim$~3900    days, and   image~B  at
HJD~$\sim$~3500 days.  We show that  the continuum  of these two  images
becomes bluer as they   get  brighter, as expected from   microlensing
magnification of an accretion disk.

We also report microlensing-induced variations of the BELs, both in
their integrated line intensities and in their profiles.  In image A, we
find that the profile of the \ion{C}{III]} line is broadened during
the brightening episode at HJD~$\sim$~3900 days.  The
\ion{C}{III]} line profile in image C seems also to be 
broadened at several epochs. Broadening of the BELs in image B is
less obvious.

Variations in the BEL intensities are detected mainly  in images A and
B. Our  measurements  suggest    that  higher ionization   BELs   like
\ion{C}{IV},  \ion{C}{III]},  are  more  magnified than lower   
ionization   lines like \ion{Mg}{II}. This  is  compatible with
reverberation mapping studies and  a stratified structure of  the BLR.
There    is  marginal evidence that   regions   of different sizes are
responsible for the \ion{Fe}{II+III} emission.

Finally, we estimate   the  differential extinction between   pairs of
quasar images due to  dust in the  lensing galaxy to  be in the range
0.1-0.3 mag, with images C  and D being  the most reddened.
This amount of differential extinction is  too small to  explain  
the  large microlensing-magnification ratios
involving images A and C.  Long-term microlensing,
beyond the duration of our observations, is therefore present
in these images.

The timescales of  the microlensing variations  in \obj\
are  such that  each  microlensing  event  lasts about  one  observing
season  (i.e. 8 months), with gaps of several months between  events. 
This means that
a relatively loose  observing  rate of 1 spectrum   every 15  days  is
sufficient to sample the events well enough. In addition, the Einstein Cross
is the  lensed quasar   with the  fastest and   sharpest  microlensing
events. It is therefore unique  in the sense that only  a few years of
monitoring can truly  constrain the quasar structure  on parsec scales
(Kochanek \cite{kochanek}).  In addition, in the case  of the Einstein Cross, the
very different behaviours of the BELs and the continuum with  res\-pect to
microlensing offer considerable hope to  reconstruct the  two types of
regions independently, using ray-shooting simulations.

With two more years of data, we expect to map a total of up to half a
dozen microlensing events in the four quasar images,
hence providing a unique and useful data set for microlensing 
and quasar studies.

\begin{acknowledgements}
We are  extremely grateful to all ESO staff for their excellent work. 
The observations  presented  in  this  article have involved   a lot of
efforts  from	the ESO staff  operating  FORS1,   to ensure
accurate and reproducible mask alignment,  to keep the best
possible   temporal sampling, and to meet the  requested
seeing value.  This project is  partially supported  by the Swiss
National Science Foundation (SNSF).
\end{acknowledgements}

\end{document}